\documentclass[pra,twocolumn,floatfix,superscriptaddress]{revtex4-1}
\usepackage{dcolumn,amsmath}
\usepackage{graphicx}
\usepackage{braket}
\usepackage{diagbox}
\usepackage{bm}
\usepackage{amssymb}
\usepackage{hyperref}
\usepackage{multirow}
\usepackage{footnote}
\usepackage{subcaption}
\usepackage{ragged2e}
\usepackage{xcolor}
\usepackage{pdfcomment}
\usepackage[justification=justified,labelfont=bf,singlelinecheck=off]{caption}
\usepackage{threeparttable,booktabs} 
\usepackage{subcaption,siunitx,booktabs}
\usepackage{caption,afterpage,tabularx}
\usepackage[tableposition=top]{caption}

\newcommand{\RUG}{
Van Swinderen Institute for Particle Physics and Gravity,
University of Groningen, Nijenborgh 4, 9747AG Groningen, The Netherlands and Nikhef, National Institute for Subatomic Physics, Science Park 105, 1098 XG Amsterdam, The Netherlands 
\\
}

\newcommand{\CU}{
Department of Physical and Theoretical Chemistry \& Laboratory for Advanced Materials, Faculty of Natural Sciences, 
Comenius University, Mlynská dolina, 841 04 Bratislava, Slovakia\\
}

\newcommand{\VU}{
Division of Theoretical Chemistry, Faculty of Sciences, 
Vrije Universiteit Amsterdam, De Boelelaan 1083, 1081 HV Amsterdam, The Netherlands\\
}

\newcommand{\VULaserLab}{
 Department of Physics and Astronomy, and LaserLaB, Vrije Universiteit Amsterdam,
De Boelelaan 1081, 1081 HV Amsterdam, The Netherlands\\
}

\begin{document}

\title{High accuracy theoretical investigations of CaF, SrF, and BaF and implications for laser-cooling}
\author{Yongliang Hao}
\affiliation{\RUG}
\author{Luka\v{s} F. Pa\v{s}teka}
\affiliation{\CU}
\author{Lucas Visscher}
\affiliation{\VU}
\author{and the NL-$e$EDM collaboration: Parul Aggarwal}
\affiliation{\RUG}
\author{Hendrick L. Bethlem}
\affiliation{\VULaserLab}
\author{Alexander Boeschoten}
\affiliation{\RUG}
\author{Anastasia Borschevsky}
\email{a.borschevsky@rug.nl}
\affiliation{\RUG}
\author{Malika~Denis}
\affiliation{\RUG}
\author{Kevin Esajas}
\affiliation{\RUG}
\author{Steven~Hoekstra}
\affiliation{\RUG}
\author{Klaus Jungmann}
\affiliation{\RUG}
\author{Virginia R. Marshall}
\affiliation{\RUG}
\author{Thomas B. Meijknecht}
\affiliation{\RUG}
\author{Maarten C. Mooij}
\affiliation{\VULaserLab}
\author{Rob~G.~E.~Timmermans}
\affiliation{\RUG}
\author{Anno Touwen}
\affiliation{\RUG}
\author{Wim Ubachs}
\affiliation{\VULaserLab}
\author{Lorenz Willmann}
\affiliation{\RUG}
\author{Yanning Yin}
\affiliation{\RUG}
\author{Artem Zapara}
\affiliation{\RUG}

\date{\today}

\begin{abstract}

The NL-eEDM collaboration is building an experimental setup to search for the permanent electric dipole moment of the electron in a slow beam of cold barium fluoride molecules [Eur. Phys. J. D, \textbf{72}, 197 (2018)]. Knowledge of molecular properties of BaF is thus needed to plan the measurements and in particular to determine the optimal laser-cooling scheme. Accurate and reliable theoretical predictions of these properties require incorporation of both high-order correlation and relativistic effects in the calculations. In this work theoretical investigations of the ground and the lowest excited states of BaF and its lighter homologues, CaF and SrF, are carried out in the framework of the relativistic Fock-space coupled cluster (FSCC) and multireference configuration interaction (MRCI) methods. Using the calculated molecular properties, we determine the Franck-Condon factors (FCFs) for the $A^2\Pi_{1/2} \rightarrow X^2\Sigma^{+}_{1/2}$ transition, which was successfully used for cooling CaF and SrF and is now considered for BaF. For all three species, the FCFs are found to be highly diagonal. Calculations are also performed for the $B^2\Sigma^{+}_{1/2} \rightarrow X^2\Sigma^{+}_{1/2}$ transition recently exploited for laser-cooling of CaF; it is shown that this transition is not suitable for laser-cooling of BaF, due to the non-diagonal nature of the FCFs in this system. Special attention is given to the properties of the $A'^2\Delta$ state, which in the case of BaF causes a leak channel, in contrast to CaF and SrF species where this state is energetically above the excited states used in laser-cooling. We also present the dipole moments of the ground and the excited states of the three molecules and the transition dipole moments (TDMs) between the different states. Finally, using the calculated FCFs and TDMs we determine that the $A^2\Pi_{1/2} \rightarrow X^2\Sigma^{+}_{1/2}$ transition is suitable for transverse cooling in BaF.


\end{abstract}

\maketitle

\section{INTRODUCTION\label{I}}

Heavy diatomic molecules are currently considered to be the most sensitive systems used in the search for the electron electric dipole moment (electron-EDM) \cite{Safronova2018}. The large effective electric field, which the valence electron in these molecules is exposed to \cite{Dzuba2012}, allows for a huge sensitivity enhancement compared to a measurement on an atom.

In the ongoing experiments on YbF \cite{Hudson2011} and ThO \cite{Baron2014,Andreev2018}, and the planned experiment on BaF \cite{Aggarwal2018}, precision measurements are performed on a beam of molecules using the Ramsey separated oscillatory fields method \cite{Ramsey1990}. In the region of the experiment where the molecular beam interacts with carefully defined electric and magnetic fields, the electron-EDM can become visible in the correlation of an energy level shift with the direction of the electric field. The sensitivity of such a measurement scales with the square root of the total number of molecules used in the experiment, and linearly with the coherent interaction time in the Ramsey detection scheme. To optimize the sensitivity, the interaction time in these experiments can be increased by reducing the longitudinal velocity of the molecular beam by using a cryogenic beam source or by Stark deceleration. However, if the transverse velocity spread of the molecular beam is not also reduced, the increase in the interaction time will be offset by an increased transverse spreading of the molecular beam during the transition of the interaction zone, and the sensitivity of the experiment will not be improved. Transverse laser-cooling of molecular beams can reduce the spread of the molecular beam to a negligible level, provided the internal structure of the molecule is suitable. This leads to an increase in the number of molecules and thereby opens the way to experiments with very long interaction times and an improved sensitivity for measuring the electron-EDM. The possibility to exert both laser-cooling and Stark deceleration on the BaF molecule makes this species a candidate for a successful measurement of the electron-EDM  \cite{Aggarwal2018}.

The prospects of laser-cooling and trapping of molecules \cite{Di2004} have led to a considerable interest in both experimental and theoretical communities. The first molecule to be laser cooled was SrF \cite{Shuman2010}, followed by YO \cite{Hummon2013}, CaF \cite{Zhelyazkova2014}, and YbF \cite{Lim2018}. Recently, laser-cooling of the first polyatomic molecule, SrOH, was demonstrated \cite{Kozyryev2017sisyphus} and has been proposed for heavier molecules, such as RaOH and YbOH, and larger polyatomic molecules like YbOCH$_3$ \cite{Isaev2017,Kozyryev2017}.

There are a number of key factors that determine whether a  given molecule is suitable for laser-cooling \cite{Di2004}.  One is having strong one-photon transitions to ensure the high photon-scattering rates needed for efficient momentum transfer. The oscillator strengths of the transitions can be determined using the transition dipole moments (TDMs) between the states. A second
requirement is a rotational structure with a closed optical cycle; this is available in $^2\Pi - ^2\Sigma^+$ and $^2\Sigma^+ - ^2\Sigma^+$ transitions. A third
condition concerns the Franck-Condon factors (FCFs) which govern the vibronic transitions between different electronic states. Highly diagonal FCFs provide a near-closed optical cycle in the vibronic structure, therewith limiting the required repumping. Finally,  there should either be no intervening electronic states to which the upper state could radiate and cause leaks in the cooling cycle, or the transitions to such states should be suppressed.

Thus, the suitability of BaF for laser-cooling depends critically on its energy level structure, lifetimes of  its excited states, vibrational branching ratios, and electronic transition probabilities. This paper aims to determine these properties at the highest possible level of computational accuracy, to conclude on the suitability of BaF for laser-cooling, and to suggest the optimal laser-cooling scheme.

We perform high-accuracy relativistic Fock-space coupled cluster (FSCC) calculations of the spectroscopic constants of BaF and its lighter homologues CaF and SrF; based on these values we provide predictions of the FCFs of the $A^2\Pi_{1/2}- X^2\Sigma^{+}_{1/2}$ laser-cooling transition, the alternative cooling transition $B^2\Sigma^{+}_{1/2}-X^2\Sigma^{+}_{1/2}$, and the possible leak transition $A^2\Pi_{1/2} - A'^2\Delta_{3/2}$. We also carry out calculations of the dipole moments (DMs) and transition dipole moments (TDMs) for the six lowest states of the selected molecules, using in this case the relativistic multireference configuration interaction method (MRCI). These are the first comprehensive relativistic high-accuracy investigations of the spectroscopic properties of these molecules. The ground and the excited state properties are treated on the same footing, and similar accuracy is expected for all the levels investigated here. 

In the following, we start in Section \ref{Previous} with a brief overview of previous experimental and theoretical studies of the three molecules. In Section \ref{Methods} the methods employed in our calculations are introduced. Section \ref{Calculations} contains our theoretical results for spectroscopic constants, Franck-Condon factors, dipole moments, transition dipole moments, and lifetimes of the excited states. The implications of these results for possible laser-cooling schemes are discussed in Section \ref{Lasercooling}. 

\section{Previous investigations}
\label{Previous}

Numerous theoretical studies of the electronic structure and other properties of BaF and its lighter homologues were carried out, using a variety of methods. The majority of these investigations were performed in a nonrelativistic framework. The main system of interest here, BaF, was recently investigated using the effective core potential (ECP) based complete active space self-consistent field approach combined with the multi-reference configuration interaction method (CASSCF+MRCI) \cite{Tohme2015}. This study provides the spectroscopic constants, the static and the transition dipole moments, and the static dipole polarizabilities of the ground and the 41 lowest doublet and quartet electronic states of this system. The drawbacks of this extensive investigation are in the rather limited size of the employed basis sets and in the fact that spin-orbit coupling is neglected altogether. Shortly after, Kang \textit{et al.} \cite{Kang2016} published a paper where a similar approach (CASSCF+MRCI) was used to investigate the properties of BaF, including the Franck-Condon factors for the transitions between the lowest states. Here,  much higher quality basis sets were used, and spin-orbit coupling (SOC) effects were included at the MRCI level. The FCFs for the transition between the low-lying states of BaF were reported by Chen \textit{et al.} \cite{Chen2016s} using the Rydberg-Klein-Rees (RKR) approach, and by Karthikeyan \textit{et al.} \cite{Karthikeyan2013} and Xu \textit{et al.} \cite{Xu2017} within the Morse potential model (MPM). The DM of the ground state of BaF was also studied using the relativistic restricted active space approach combined with configuration interaction method (RASCI) \cite{Nayak2006}, by relativistic coupled cluster method (RCCSD/RCCSD(T)) \cite{Prasannaa2016,Abe2018,Fazil2018}, and using relativistic effective core potential approach based on the restricted active space self-consistent-field theory (AREP-RASSCF) \cite{Kozlov1997}.

Earlier, in the work of Westin \textit{et al.} \cite{Westin1988}, the transition energies between low-lying electronic states of BaF were obtained based on density functional theory (DFT) method. The spectroscopic constants ($\alpha_e$ and $\omega_e\chi_e$) and dipole moments of the ground state of BaF and its homologues were calculated by T\"orring \textit{et al.} \cite{Torring1984} using the ionic Rittner model \cite{Rittner1951}. Subsequently, these authors applied an electrostatic polarisation model (EPM) \cite{Torring1989} to evaluate the energies and the dipole moments of the low-lying excited states of alkaline earth metal monohalides, including BaF. The transition energies as well as the DMs and TDMs of the lowest excited states of CaF, SrF, and BaF were reported by Allouche \textit{et al.} \cite{Allouche1993} using the Ligand Field Method (LFM), where model potential functions are used to describe the electronic structure of alkaline earth metal ions.

The majority of theoretical investigations on CaF and SrF were carried out using the configuration interaction approach, either within its single reference (CISD) \cite{Langhoff1986} or multireference variant \cite{Bundgen1991,Pelegrini2005,Yang2007}. Most recently, two studies were published, presenting the spectroscopic constants and the DMs of the two molecules obtained both by the CASSCF+MRCI approach  and using the second-order multireference Rayleigh-Schr\"{o}dinger perturbation theory (CASSCF+RSPT2) \cite{Jardali2014,El2017}. Some single reference coupled cluster studies are also available \cite{Harrison2002,Prasannaa2014,Sasmal2015,Kosicki2017,Hou2018,Abe2018,Fazil2018}. Other approaches used for calculations of the spectroscopic constants, the DMs, and the TDMs of CaF and SrF are the ligand field method \cite{Rice1985,Allouche1993}, the electrostatic polarisation model \cite{Torring1989,Mestdagh1991}, the finite-difference Hartree-Fock (FDHF) approach \cite{Kobus2000}, the second order M\o ller-Plesset perturbation theory (MP2) \cite{Buckingham1993}, the effective one-electron variational eigenchannel R-matrix method (EOVERM)  \cite{Raouafi2001}, and the ionic model \cite{Torring1984}. Barry \cite{Shuman2010,Barry2013} obtained the potential energy curves of SrF using experimental spectroscopic constants within the first-order Rydberg-Klein-Rees (RKR) approach and subsequently evaluated the FCFs for the transition $A^2\Pi_{1/2} \rightarrow X^2\Sigma^{+}_{1/2}$.

Many spectroscopic constants of the ground and the lowest excited states of BaF were determined experimentally with high precision \cite{NIST,Huber1979,Ryzlewicz1980,Barrow1988,Bernard1992}, along with the DM of its ground state \cite{Ernst1986} and electronic transition dipole moments between its lowest levels \cite{Berg1993,Berg1998}, which were extracted from the measured lifetimes using calculated FCFs. There is also a significant amount of experimental data available on the properties of its lighter homologues, CaF and SrF. Here, we cite the most recent and precise values available: Refs. \cite{NIST,Barrow1967,Dagdigian1974,Field1975,Steimle1977,Nakagawa1978,Ernst1983,Huber1979,
Verges1993,steimle1993molecular,Colarusso1996,Kaledin1999,Nitsch1988,Sheridan2009} 
for the spectroscopic constants, Refs. \cite{Childs1984, Ernst1985, Ernst1989, Kandler1989} for the static and transition dipole moments,  Refs. \cite{Dagdigian1974,Berg1996} for the lifetimes, and a single measurement of the FCFs of the $A-X(0-0)$ transition in CaF \cite{Wall2008}.

RaF, the heavier homologue of BaF, was also proposed for laser-cooling and for use in experiments to search for physics beyond the Standard Model. Its spectroscopic properties were investigated within the relativistic Hartree-Fock and the density functional theory methods \cite{Isaev2010,Isaev2012,Borschevsky2012,Rorschevsky2013,Isaev2014}, and using the relativistic coupled cluster approach \cite{KudPetSkr14}. This molecule, along with the lighter BeF and MgF, is, however, outside the scope of the present work.

\section{METHODS\label{Methods}}
Relativistic effects can have a significant influence on atomic and molecular properties \cite{Pyykko1988}, in particular in case of heavier atoms and molecules, represented by BaF in this study. Thus, we have carried out all the calculations within the relativistic framework, using the DIRAC15 program package \cite{DIRAC15}. In order to conserve computational effort, we have replaced the traditional 4-component Dirac-Coulomb (DC) Hamiltonian by the exact 2-component Hamiltonian (X2C) \cite{Iliavs2007,Saue2011}. This approach allows a significant decrease in computational time and expense, while reproducing very well the results obtained using the 4-component DC Hamiltonian, as tested for a variety of species and properties \cite{Bast2009,Sikkema2009,Knecht2010}. In this work, we have used the molecular mean-field implementation of the approach, X2Cmmf \cite{Sikkema2009} and included the Gaunt interaction \cite{Gaunt1929}. This interaction is part of the Breit term, which corrects the 2-electron part of the Dirac-Coulomb Hamiltonian up to the order of $(Z \alpha)^2$ \cite{Breit1929}. The Breit correction was shown to be of importance even for light molecules \cite{Visser1992}; we thus include the Gaunt term in our calculations, for achieving optimal accuracy (the full Breit term is to date not implemented in the DIRAC program). All the calculations were performed for the $^{138}$BaF, $^{88}$SrF, and $^{40}$CaF isotopologues. 

In order to obtain the spectroscopic constants of the ground and excited states of the molecules and the Franck-Condon factors for transitions between these states, we have calculated the potential energy curves using the multireference relativistic Fock space coupled cluster approach \cite{Eliav1994}. FSCC is considered one of the most powerful methods for high-accuracy calculations of atomic and molecular properties of small heavy species and it is particularly well suited for treating excited states \cite{Eliav2017}. Within the framework of this approach an effective Hamiltonian ($H_{\rm eff}$) is defined and calculated in a low-dimensional model ($P$) space, constructed from zero-order wave functions (Slater determinants), with eigenvalues approximating some desirable eigenvalues of the physical Hamiltonian. The effective Hamiltonian has the form \cite{Lindgren1982}
\begin{eqnarray}
H_{\rm eff}=PH\Omega P,
\label{Heff}
\end{eqnarray}
where $\Omega$ is the normal-ordered wave operator,
\begin{eqnarray}
\Omega={\exp(S)}.
\label{Omega}
\end{eqnarray}
The excitation operator $S$ is defined with respect to a closed-shell reference determinant (vacuum state), and partitioned according to the number of valence holes ($m$) and valence electrons ($n$) with respect to this reference:
\begin{eqnarray}
S= \sum_{m \geqslant 0} \sum_{n \geqslant 0} \left( \sum_{l \geqslant m+n} S_{l} ^ {(m,n)} \right) .
\label{S}
\end{eqnarray} 
Here $l$ is the number of excited electrons. Current implementation of the relativistic FSCC method \cite{Eliav1994} is limited to $l \leqslant 2$, corresponding to single and double excitations, and thus, $m+n \leqslant 2$, which in practice means that we are able to treat atoms and molecules with up to two valence electrons or holes.
 
\begin{figure*}[t]
\centering
\includegraphics[scale=0.99,width=0.99\linewidth]{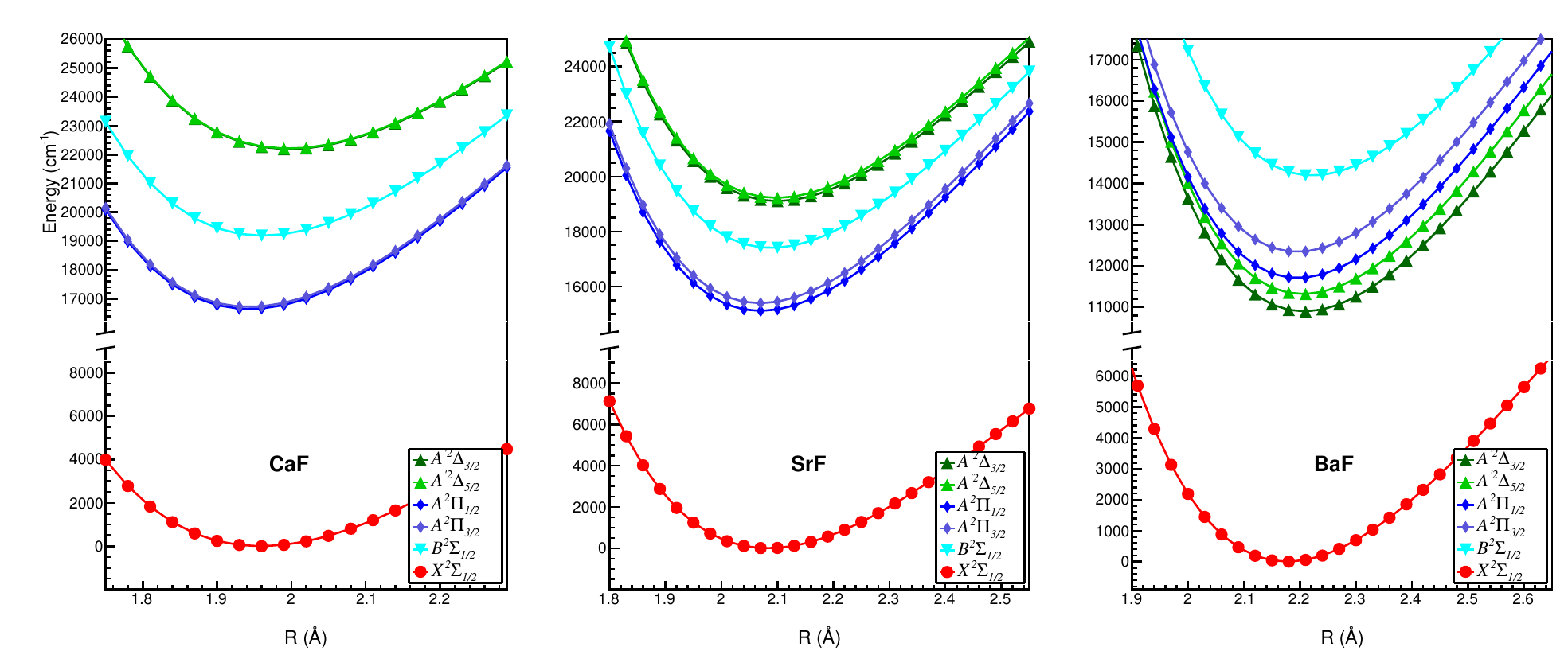}
\justifying
\caption{Potential energy curves for the low-lying states of CaF, SrF, and BaF (Color online).}
\label{PEC}
\end{figure*}

\begin{table*}[t]
\caption{Spectroscopic constants of the ground and the low-lying excited states of CaF.\label{CaF-SP}
}
\begin{ruledtabular}
\begin{tabular}{lllllllll}
& $X^2{\Sigma^{+}_{1/2}}$ & $A'^2{\Delta}_{3/2}$ & $A'^2{\Delta}_{5/2}$ & $A^2{\Pi}_{1/2}$ & $A^2{\Pi}_{3/2}$ & $B^2{\Sigma^{+}_{1/2}}$ & Method & Reference\\
\hline
$R_{e}$(${\AA}$)                  &1.958&1.997&1.996&1.943&1.943&1.961&X2C-FSCC&This work\\
                                  
                                  &1.965&&&&&&CISD&\cite{Langhoff1986}\\
                                  &1.975&1.998&1.998&1.957&1.957&1.977&MRCI$^a$&\cite{Bundgen1991}\\
                                  &1.971&&&1.954&1.954&&MRCI$^a$&\cite{Pelegrini2005}\\   
                                  &2.001&2.027&2.027&1.981&1.981&1.992& MRCI$^a$&\cite{Yang2007}\\
                                  &2.015&2.050&2.050&2.001&2.001&2.009&CASSCF+MRCI$^a$&\cite{El2017}\\
                                  &2.015&2.071&2.071&2.008&2.008&2.043&CASSCF+RSPT2$^a$&\cite{El2017}\\
                                  &1.967&&&1.952&1.952&&Experiment$^a$&\cite{NIST,Huber1979}\\
                                  &&1.993(3)&1.993(3)&&&&Experiment$^a$&\cite{Verges1993}\\
                                  &&&&1.9374(1)&1.9374(1)&1.9555(3)&Experiment$^a$&\cite{Kaledin1999}\\\hline

${\omega}_{e}$(cm$^{-1}$)         &586.2&529.4&529.5&594.6&594.6&572.8&X2C-FSCC&This work\\
                                  
                                  &587&&&&&&CISD&\cite{Langhoff1986}\\                           
                                  & 581.2&558.9&558.9&579.9&579.9&551.5&MRCI$^a$&\cite{Bundgen1991}\\
                                  & 612.5&&&624.0&624.0&&MRCI$^a$&\cite{Pelegrini2005}\\   
                                  &572.4& 506.1& 506.1&578.6&578.6&571.4&MRCI$^a$&\cite{Yang2007}\\
                                  &524.3&498.2&498.2&563.4&563.4&512.6&CASSCF+MRCI$^a$&\cite{El2017}\\
                                  &518.6&462.4&462.4&511.4&511.4&472.5&CASCF+RSPT2$^a$&\cite{El2017}\\      
                                  &581.1(9)&&&586.8(9)&&&Experiment$^a$&\cite{Field1975}\\
                                  &&528.57(1)&528.57(1)&&&&Experiment$^a$&\cite{Verges1993}\\  &&&&594.513(50)&594.513(50)&572.424(80)&Experiment$^a$&\cite{Kaledin1999}\\\hline

${\omega}_{e}{\chi}_{e}$(cm$^{-1}$)&2.90&2.86&2.85&3.03&3.04&3.13&X2C-FSCC&This work\\
                                  
                                  &3.70&&&3.77&3.77&&MRCI$^a$&\cite{Pelegrini2005}\\   
                                  &2.65&2.75&2.75&2.60&2.60&3.24&MRCI$^a$&\cite{Yang2007}\\
                                  &&2.77(9)&2.77(9)&&&&Experiment$^a$&\cite{Verges1993}\\
                                  &&&&3.031(20)&3.031(20)&3.101(37)&Experiment$^a$&\cite{Kaledin1999}\\\hline                                          
${B}_{e}$(cm$^{-1}$)              &0.341&0.328&0.329&0.347&0.347&0.341&X2C-FSCC&This work\\
                                  
                                  &0.335&0.328&0.328&0.342&0.342&0.335&MRCI$^a$&\cite{Bundgen1991}\\
                                  &0.327&0.319&0.319&0.334&0.334&0.330&MRCI$^a$&\cite{Yang2007}\\
                                  &0.322&0.311&0.311&0.326&0.326&0.324&CASSCF+MRCI$^a$&\cite{El2017} \\
                                  &0.322&0.305&0.305&0.324&0.324&0.313&CASSCF+RSPT2$^a$&\cite{El2017}\\      
                                  &0.343704(23)&&&0.348744(27)&0.348744(27)&&Experiment&\cite{Nakagawa1978}\\
                                  &&0.3295&0.3295&&&&Experiment$^a$&\cite{Verges1993}\\ &&&&0.348781(5)&0.348781(5)&0.342345(10)&Experiment$^a$&\cite{Kaledin1999}\\\hline
                                  
$T_{e}$(cm$^{-1}$)                &0&22187&22207&16647&16720&19191&X2C-FSCC&This work\\
                                  &&24950&24950&17998&17998&22376&LFM&\cite{Rice1985}\\ 
                                  &&17690&17690&16340&16340&18620&EPM&\cite{Torring1989}\\
                                  &&24851&24851&17712&17712&20069&MRCI$^a$&\cite{Bundgen1991}\\
                                  &&22552&22552&18217&18217&21486&LFM&\cite{Allouche1993}\\
                                  &&&&16421&16421&&MRCI$^a$&\cite{Pelegrini2005}\\   
                                  &&20697&20697&15627&15627&19512&MRCI(CBS)$^a$&\cite{Yang2007}\\
                                  &&22113&22113&16544&16544&19013&CASSCF+MRCI$^a$&\cite{El2017}\\
                                  &&25337&25337&16574&16574&21016&CASSCF+RSPT2$^a$&\cite{El2017}\\  
                                  &&21567.76(1)&21580.10(1)&&&&Experiment&\cite{Verges1993}\\
                                  &&&&16491.036(50)&16562.465(50)&18840.190(60)&Experiment&\cite{Kaledin1999}\\                                
\end{tabular}
\end{ruledtabular}
  \begin{tablenotes}
      \footnotesize
      \item[\emph{a}]{$^a$ As this study neglects spin-orbit coupling, the same values of the spectroscopic constants are given for the $A'^2{\Delta}_{3/2}$ and $A'^2{\Delta}_{5/2}$ and the $A^2{\Pi}_{1/2}$ and $A^2{\Pi}_{3/2}$ states.}\\
  \end{tablenotes}
\end{table*}
 
The molecules of interest all have a single valence electron and a $ ^2 \Sigma^{+}_{1/2}$ ground state configuration. We thus start our calculations from the closed-shell positively charged ions, CaF$^+$, SrF$^+$, and BaF$^+$. After solving the coupled cluster equations for these closed-shell reference ions, we proceed to add an electron to reach the neutral states, for which additional CC equations are solved to obtain the correlated ground and excited state energies. In this work, we were interested in the $X^2 \Sigma^{+}_{1/2}$, $A^2 \Pi_{1/2}$, $A^2 \Pi_{3/2}$, $A'^2 \Delta_{3/2}$, $A'^2 \Delta_{5/2}$, and $B^2 \Sigma^{+}_{1/2}$ states. We have thus defined the model space $P$ to contain the appropriate $\sigma$, $\pi$, and $\delta$ orbitals.

In order to reach optimal accuracy very large basis sets were used in the calculations, and higher angular momentum basis functions were added manually to the available sets. For all the elements involved in our calculation, we have employed the relativistic basis sets of Dyall \cite{Dyall2009,Dyall2016}. Singly augmented pVQZ basis set (s-aug-pVQZ) was used for fluorine; for Sr and Ba we used the doubly augmented pVQZ basis sets (d-aug-pVQZ), to which we manually added two $h$-type functions with exponent values of 0.48 and 0.25. For CaF the pVQZ did not provide sufficient quality for description of the $\Delta$ states, while on the other hand the $h$-type functions had very little effect on the calculated transition energies. We thus used the doubly augmented core-valence CVQZ basis set for this element (this basis set has two additional $d$, one additional $f$ and one additional $g$ functions compared to the d-aug-pVQZ basis). Convergence of the obtained spectroscopic constants (in particular excitation energies) with respect to the basis set size was verified. We have correlated 34 electrons in case of BaF and SrF and 24 electrons for CaF.
 
After obtaining the potential energy curves, we have used the Dunham \cite{Dunham1932} 
programme (written by V. Kell\"{o} of the Comenius University, \cite{Kel18}) to calculate the spectroscopic constants: the equilibrium bond lengths ($R_e$), the harmonic and anharmonic vibrational frequencies (${\omega}_e$ and ${\omega}_e{\chi}_e$), the adiabatic transition energies ($T_e$), and the rotational constants ($B_e$). The Frank-Condon factors between the low lying vibrational levels of the ground state and the excited states were extracted using the LEVEL16 program of Le Roy \cite{le2017level}. 

The calculations of the dipole moments and the transition dipole moments were carried out using the MRCISD method \cite{Shavitt1977} as implemented in the LUCIAREL module \cite{FleOlsVis03,KneJenFle10} of the DIRAC15 program package \cite{DIRAC15}. The change of method is needed because calculation of TDMs is not yet implemented on the coupled cluster level. Since the MF (M=Ca, Sr, Ba) molecule is considerably ionic \cite{Knight1971}, in first approximation we can describe this system as a metal cation M$^+$ perturbed by the presence of the F$^-$ anion \cite{Jardali2014}. Hence, the valence electronic structure of MF is qualitatively similar to M$^+$: $ns^1$. All the excited states of interest can be similarly described by the single unpaired valence electron being excited into the low lying empty valence $d$ shell of the M$^+$ cation.
The configuration space was thus defined as one electron spanning the 6 orbitals corresponding to the metal atomic orbitals: $ns$ and $(n-1)d$, thus describing two $^2\Sigma$, one $^2\Pi$ and one $^2\Delta$ state. In order to describe the orbitals equally well for all states we used the average-of-configuration DHF reference orbitals \cite{Desclaux1975} with one electron occupying the same 6 orbitals as were included in the configuration space. The correlation space extended down to the $(n-1)$ shell of the M$^+$ cation and $2s$, $2p$ orbitals of F$^-$ anion (i.e. 8 additional occupied orbitals) and virtual orbitals with energies over 10 a.u were cut off. 
For the DMs, both computational methods are appropriate; there we use the FSCC values to test the performance and the validity of MRCI for the TDM calculations. The same basis sets were employed as for the calculations of the potential energy curves.

\section{RESULTS AND DISCUSSION\label{Calculations}}

\subsection{Potential Energy Curves}

The calculated potential energy curves of the ground and the low-lying excited states of the three molecules are shown in Fig.~\ref{PEC}. As expected, the energy splitting between the $\Omega$ resolved states tends to be larger the heavier the molecule becomes due to the relativistic effects playing a more important role in heavier species. An important difference in the electronic structure of the three molecules is in the location of the $A'^2 \Delta$ states. For CaF and SrF, these states are higher than the $A^2 \Pi$ states, even higher than the $B^2 \Sigma^{+}$ state, while for BaF they are lower in energy and transitions to the $A'^2 \Delta_{3/2}$ state could constitute a leak in the cooling cycle.

\subsection{Spectroscopic Constants}

Tables ~\ref{CaF-SP}, ~\ref{SrF-SP}, and ~\ref{BaF-SP} contain the calculated spectroscopic constants of the three molecules, along with experimental values where available and earlier theoretical results. Throughout this paper, all the molecular constants are defined in the usual way \cite{herzberg2013molecular}. Overall, our calculations are in excellent agreement with experiment. For most of the values, the error is less than 1\%; the largest relative error (of a few percent) is for the anharmonicity correction ${\omega}_e{\chi}_e$. However, for these constants the experimental uncertainty is often rather high. The calculated transition energies are generally slightly overestimated due to the neglect of the triple excitations, which are to date not implemented in the FSCC approach.

The present results can be compared to the most recent theoretical investigations. For CaF, these are the nonrelativistic MRCI calculations of Ref. \cite{Yang2007} and the nonrelativistic CASSCF+MRCI and CASSCF+RSPT2 values of Ref. \cite{El2017}. In both the previous studies the deviations from experiment were larger than here; in case of Ref.  \cite{El2017} use of a limited basis set led to errors on the order of 2-10\%, with MRCI performing better than the RSPT2 approach (MRCI transition energies reproduced the experiment quite well).  In Ref. \cite{Yang2007} the results were extrapolated to the complete basis set limit, resulting in lower errors of 2-5\%.

\begin{table*}[t]
\caption{Spectroscopic constants of the ground and the low-lying excited states of SrF.\label{SrF-SP} 
}
\begin{ruledtabular}
\begin{tabular}{lllllllll}
& $X^2{\Sigma^{+}_{1/2}}$ & $A'^2{\Delta}_{3/2}$ & $A'^2{\Delta}_{5/2}$ & $A^2{\Pi}_{1/2}$ & $A^2{\Pi}_{3/2}$ & $B^2{\Sigma^{+}_{1/2}}$& Method & Reference\\
\hline
$R_{e}$(${\AA}$)                  &2.083&2.099&2.098&2.069&2.069&2.089&X2C-FSCC&This work\\
                                  
                                  &2.085&&&&&&CISD&\cite{Langhoff1986}\\
                                  &2.137&2.147&2.147&2.116&2.116&2.130&CASSCF+MRCI$^a$&\cite{Jardali2014}\\
                                  &2.124&2.145&2.145&2.097&2.097&2.138&CASSCF+RSPT2$^a$&\cite{Jardali2014}\\
                                  &2.081&&&&&&CCSD(T)&\cite{Kosicki2017}\\
                                  &&&&&&2.080&Experiment&\cite{NIST,Huber1979}\\
                                  &2.0757(5)&&&&&&Experiment&\cite{Barrow1967}\\\hline
                                  
${\omega}_{e}$(cm$^{-1}$)         &500.1&475.9&476.7&508.4&508.8&492.2&X2C-FSCC&This work\\
                                  
                                  &507&&&&&&CISD&\cite{Langhoff1986}\\
                                  &475.0&454.7&454.7&491.9&491.9&480.2&CASSCF+MRCI$^a$&\cite{Jardali2014}\\
                                  &477.8&449.6&449.6&510.3&510.3&516.6& CASSCF+RSPT2$^a$&\cite{Jardali2014}\\
                                  &500.25&&&&&&CCSD(T)&\cite{Kosicki2017}\\
                                  &502.4(7)&&&&&495.8(7)&Experiment&\cite{Steimle1977}\\
                                  &501.96496(13)&&&&&&Experiment&\cite{Colarusso1996}\\\hline
                                    
${\omega}_{e}{\chi}_{e}$(cm$^{-1}$)&2.45&2.44&2.41&2.46&2.52&2.16&X2C-FSCC&This work\\
                                  
                                  &2.27(21)&&&&&2.34(21)&Experiment&\cite{Steimle1977}\\
                                  &2.204617(37)&&&&&&Experiment&\cite{Colarusso1996}\\\hline
                 
${B}_{e}$(cm$^{-1}$)              &0.249&0.245&0.245&0.252&0.252&0.247&X2C-FSCC&This work\\
                                  
                                  &0.236&0.234&0.234&0.241&0.241&0.238&CASSCF+MRCI$^a$&\cite{Jardali2014}\\
                                  &0.239&0.235&0.235&0.245&0.245&0.236&CASSCF+RSPT2$^a$&\cite{Jardali2014}\\   
                                  &0.248&&&&&&CCSD(T) & \cite{Kosicki2017}\\
                                  &0.24975935(23)&&&0.2528335(37)&0.2528335(37)&&Experiment&\cite{Sheridan2009}\\ &0.25053456(34)&&&&&0.2494103(21)&Experiment&\cite{Ernst1983}\\
                                  &&&&0.25284(3)&0.25284(3)&&Experiment$^a$&\cite{steimle1993molecular}\\
                                  &0.250534383(25)&&&&&&Experiment&\cite{Colarusso1996}\\\hline
                                  
$T_{e}$(cm$^{-1}$)                &0&19108&19225&15113&15392&17405&X2C-FSCC&This work\\
                                  
                                  &&19830&19830&15300&15300&16950&EPM$^a$&\cite{Torring1989}\\
                                  &&20553&20553&16531&16531&19295&LFM$^a$&\cite{Allouche1993}\\
                                  &&20559&20559&14506&14506&18673&CASSCF+RSPT2$^a$&\cite{Jardali2014}\\
                                  &&20790&20790&16503&16503&19005&CASSCF+RSPT2$^a$ &\cite{Jardali2014}\\
                                  &&&&&&17264.1446(12)&Experiment&\cite{Nitsch1988}\\
                                  &&&&15075.6122(7)&15357.0736(7)&&Experiment&\cite{Sheridan2009}\\
\end{tabular}
\end{ruledtabular}
 \begin{tablenotes}
     \footnotesize
      \item[\emph{a}]{$^a$ As this study neglects spin-orbit coupling, the same values of the spectroscopic constants are given for the $A'^2{\Delta}_{3/2}$ and $A'^2{\Delta}_{5/2}$ and the $A^2{\Pi}_{1/2}$ and $A^2{\Pi}_{3/2}$ states.}
 \end{tablenotes}
\end{table*}

  In case of SrF the nonrelativistic CASSCF+MRCI and CASSCF+RSPT2 \cite{Jardali2014} methods perform on a similar level, and generally somewhat better than for CaF (overall errors of 2-6\%). However, here the errors in excitation energies are larger, due to the small basis set, which is probably insufficient for an adequate description of Sr. In SrF, relativistic effects start coming into play: the spin-orbit splitting of the $A^2{\Pi}$ state is almost 300 cm$^{-1}$ and therefore, in order to achieve optimal accuracy, including spin-orbit effects is important. The ground state of SrF was also studied by the CCSD(T) approach \cite{Kosicki2017}. As expected, these results are in excellent agreement with the present values. To the best of our knowledge, no experimental information is available for the 
$A'^2{\Delta}$ states of SrF; the high-accuracy of our results for the other levels in this system supports our predictions of the properties of these states.

\begin{table*}[t]
\caption{Spectroscopic constants of the ground and the low-lying excited states of BaF.\label{BaF-SP}
}
\begin{ruledtabular}
\begin{tabular}{lllllllll}
& $X^2{\Sigma^{+}_{1/2}}$ & $A'^2{\Delta}_{3/2}$ & $A'^2{\Delta}_{5/2}$ & $A^2{\Pi}_{1/2}$ & $A^2{\Pi}_{3/2}$ & $B^2{\Sigma^{+}_{1/2}}$ &Method& Reference\\
\hline
$R_{e}$({\AA})                &2.177 & 2.207  &2.205 & 2.196 & 2.195 & 2.222 & X2C-FSCC & This work\\ 
                             
                              & 2.204 & 2.229 & 2.229 & 2.197 & 2.197 & 2.234 & CASSCF+MRCI$â$ & \cite{Tohme2015}\\
                              & 2.171 & 2.187 & 2.192 & 2.199 & 2.217 & 2.226 & CASSCF+MRCI+SOC & \cite{Kang2016}\\
                              &&&& 2.183 & 2.183 & 2.208 & Experiment$^a$ & \cite{NIST,Huber1979}\\
                              & 2.1592964(75) &&&&&& Experiment & \cite{Ryzlewicz1980}\\ \hline
                              
${\omega}_{e}$(cm$^{-1}$)     & 468.4 & 437.3 & 439.1 & 440.9 & 440.5 & 425.5 & X2C-FSCC & This work\\
                              
                              & 459.3 & 438.3 & 438.3 & 452.7 & 452.7 & 437.0 & CASSCF+MRCI$^a$ & \cite{Tohme2015}\\
                              & 474.1 & 446.3 & 423.3 & 456.7 & 417.7 & 421.7 & CASSCF+MRCI+SOC & \cite{Kang2016}\\
                              & 469.4 & 436.9 & 438.9 & 435.5 & 436.7 & 424.7 & Experiment & \cite{Barrow1988}\\ \hline
                                  
${\omega}_{e}{\chi}_{e}$(cm$^{-1}$)
                              & 1.83 & 1.84 & 1.82 & 1.90 & 1.87 & 1.81 & X2C-FSCC & This work\\
                              
                              & 1.90 & 2.02 & 1.32 & 2.55 & 1.86 & 1.83 & CASSCF+MRCI+SOC & \cite{Kang2016}\\
                              & 1.79 &&&1.68&1.82&1.88& Experiment$^b$& \cite{NIST,Huber1979}\\
                              & 1.83727(76)  &1.833(27) &1.833(27) & 1.854(12) &1.854(12) & 1.8524(37)   & Experiment$^a$ & \cite{Bernard1992}\\ \hline                             
                                  
${B}_{e}$(cm$^{-1}$)          & 0.213 & 0.207 & 0.208 & 0.209 & 0.210 & 0.205 & X2C-FSCC & This work\\
                              
                              & 0.208 & 0.203 & 0.203 & 0.209 & 0.209 & 0.202 & CASSCF+MRCI$^a$ & \cite{Tohme2015}\\
                              & 0.214 & 0.211 & 0.210 & 0.209 & 0.205 & 0.204 & CASSCF+MRCI+SOC &\cite{Kang2016}\\
                              & 0.2165297 & 0.20975 & 0.21044 &&& 0.20784 & Experiment & \cite{Barrow1988}\\ \hline

$T_{e}$(cm$^{-1}$)            & 0 & 10896 & 11316 & 11708 & 12341 & 14191 & X2C-FSCC & This work\\
                              
                              && 7420 & 7420 & 9437 & 9437 & 12663 & DFT$^a$ & \cite{Westin1988}\\
                              && 11100 & 11100 & 12330 & 12330 & 14250 & EPM$^a$ & \cite{Torring1989}\\
                              && 11310 & 11310 & 11678 & 11678 & 13381 & LFM$^a$ & \cite{Allouche1993}\\
                              && 12984 & 12984 & 11601 & 11601 & 13794& CASSCF+MRCI$^a$ & \cite{Tohme2015}\\
                              && 11582 & 12189 & 12329 & 14507 & 14022 & CASSCF+MRCI+SOC & \cite{Kang2016}\\
                              && 10734 & 11145 & 11647 & 12278 & 14063& Experiment & \cite{Barrow1988}\\ 
\end{tabular}
\end{ruledtabular}
  \begin{tablenotes}
      \footnotesize
      \item[\emph{a}]{$^a$ As this study neglects spin-orbit coupling, the same values of the spectroscopic constants are given for the $A'^2{\Delta}_{3/2}$ and $A'^2{\Delta}_{5/2}$ and the $A^2{\Pi}_{1/2}$ and $A^2{\Pi}_{3/2}$ states.}
      \item[\emph{a}]{$^b$ The experimental values for the ${\omega}_{e}{\chi}_{e}$ constants of the  $A^2{\Pi}_{1/2}$ and the $A^2{\Pi}_{3/2}$ from Ref. \cite{NIST,Huber1979} show a surprisingly large difference (0.14 cm$^{-1}$). Our results do not support this difference, and further study is needed.}
  \end{tablenotes}
\end{table*}

In BaF the order of the excited states is different to that in its lighter homologues, and the $A'^2{\Delta}$ states are below the $A^2{\Pi}$ levels. It is thus important to have high-accuracy predictions of their properties in order to estimate whether they will present a challenge in the cooling scheme. The spin-orbit splitting of the $A^2{\Pi}$ state is around 630 cm$^{-1}$ and of that of $A'^2{\Delta}$ is around 420 cm$^{-1}$. Our results reproduce very well the level ordering, the magnitude of the fine-structure splitting, and the absolute positions of the different levels as obtained from experiment. The two recent theoretical investigations of BaF used the CASSCF+MRCI approach \cite{Tohme2015,Kang2016}. The results of Ref. \cite{Tohme2015} show the $A^2{\Pi}$ states below the $A'^2{\Delta}$, most likely due to the basis set limitations. In Ref. \cite{Kang2016} a larger basis set was used, and the correct ordering of the states was reproduced. This work also included spin orbit coupling contributions, but their effect seems to be greatly overestimated, in particular for the $A^2{\Pi}$ state, where the calculated splitting is over 2000 cm$^{-1}$.

The good performance of the relativistic FSCC approach for the spectroscopic constants implies that this method is also successful in reproducing the shape of the potential energy curves (Fig.~\ref{PEC}). Therefore, we expect high-accuracy for the Frank-Condon factors presented in the next section.

\subsection{Frank-Condon Factors}\label{sec:FCFs}
In this work, we employ the extensively used r-centroid approximation \cite{Nicholls1956} for analyzing the transition rates (see e.g. Ref. ~\cite{Mccallum1979}). It factorises the transition integrals into electronic transition dipole moments and Franck-Condon factors representing the vibrational wave function overlap. Franck-Condon factors are an important parameter needed for determining whether a given system is suitable for laser-cooling. Highly diagonal FCFs would allow to limit the number of required lasers \cite{Di2004,Nguyen2011}. Therefore, we use the potential energy curves presented in the previous section to calculate the FCFs of the three molecules; the results are shown in Table~\ref{BaF-FCF-A12} for the $A^2\Pi_{1/2}-X ^2\Sigma^{+}_{1/2}$ transition, in Table~\ref{BaF-FCF2} for the $A^2\Pi_{1/2}-A'^2\Delta_{3/2}$ transition, in Table~\ref{BaF-FCF-ApX} for the $A'^2\Delta_{3/2}-X ^2\Sigma_{1/2}$ transition,  and in Table~\ref{BaF-FCF3} for the $B^2\Sigma^{+}_{1/2}-X ^2\Sigma^{+}_{1/2}$ transition. For completeness sake, we also include the FCFs of the $A^2\Pi_{3/2}-X ^2\Sigma^{+}_{1/2}$ transition in the Appendix (Table~\ref{FCF_P32_XS12}).

\begin{figure*}[t]
\centering
\includegraphics[scale=0.75,width=0.32\linewidth]{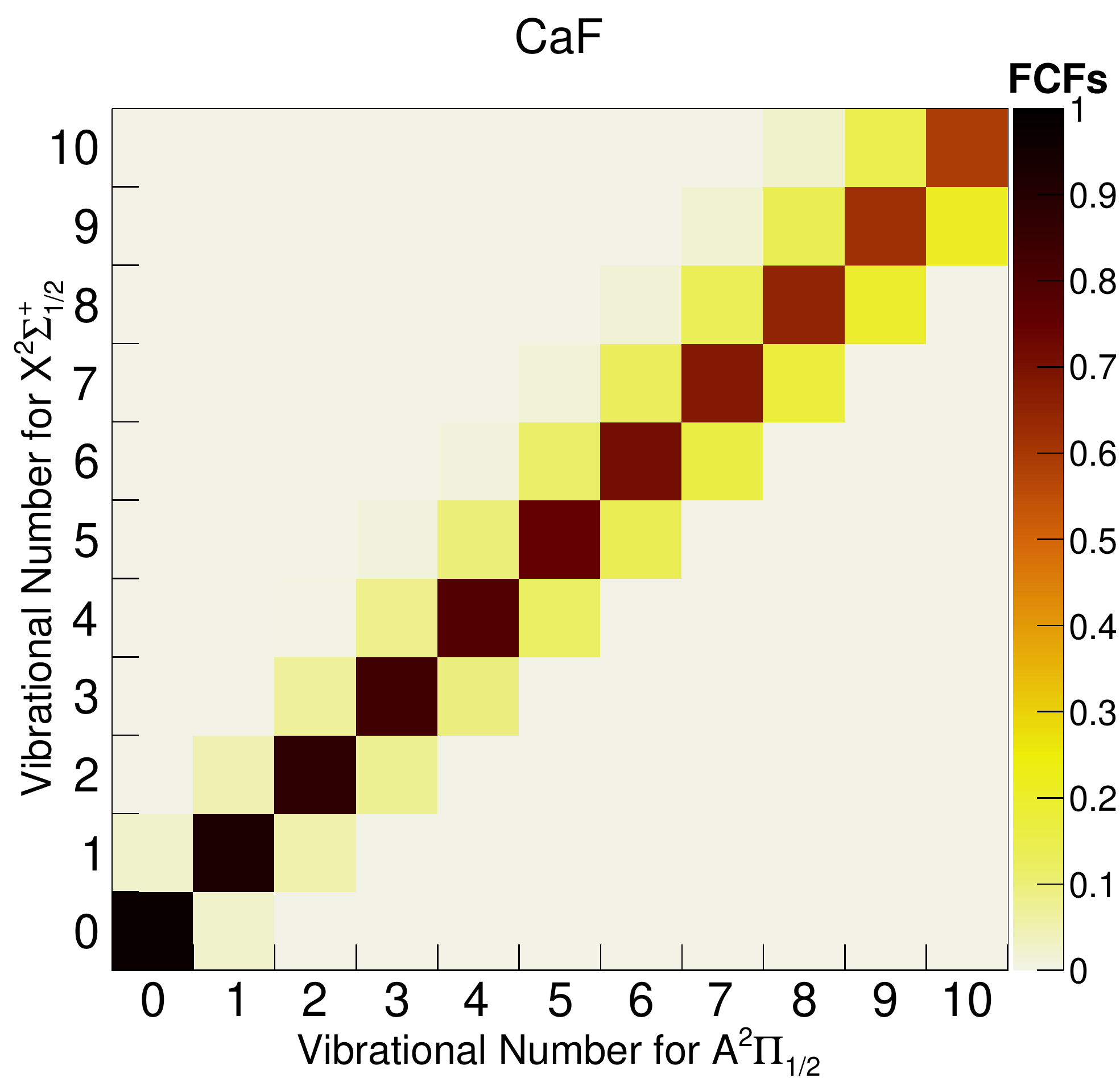}
\includegraphics[scale=0.75,width=0.32\linewidth]{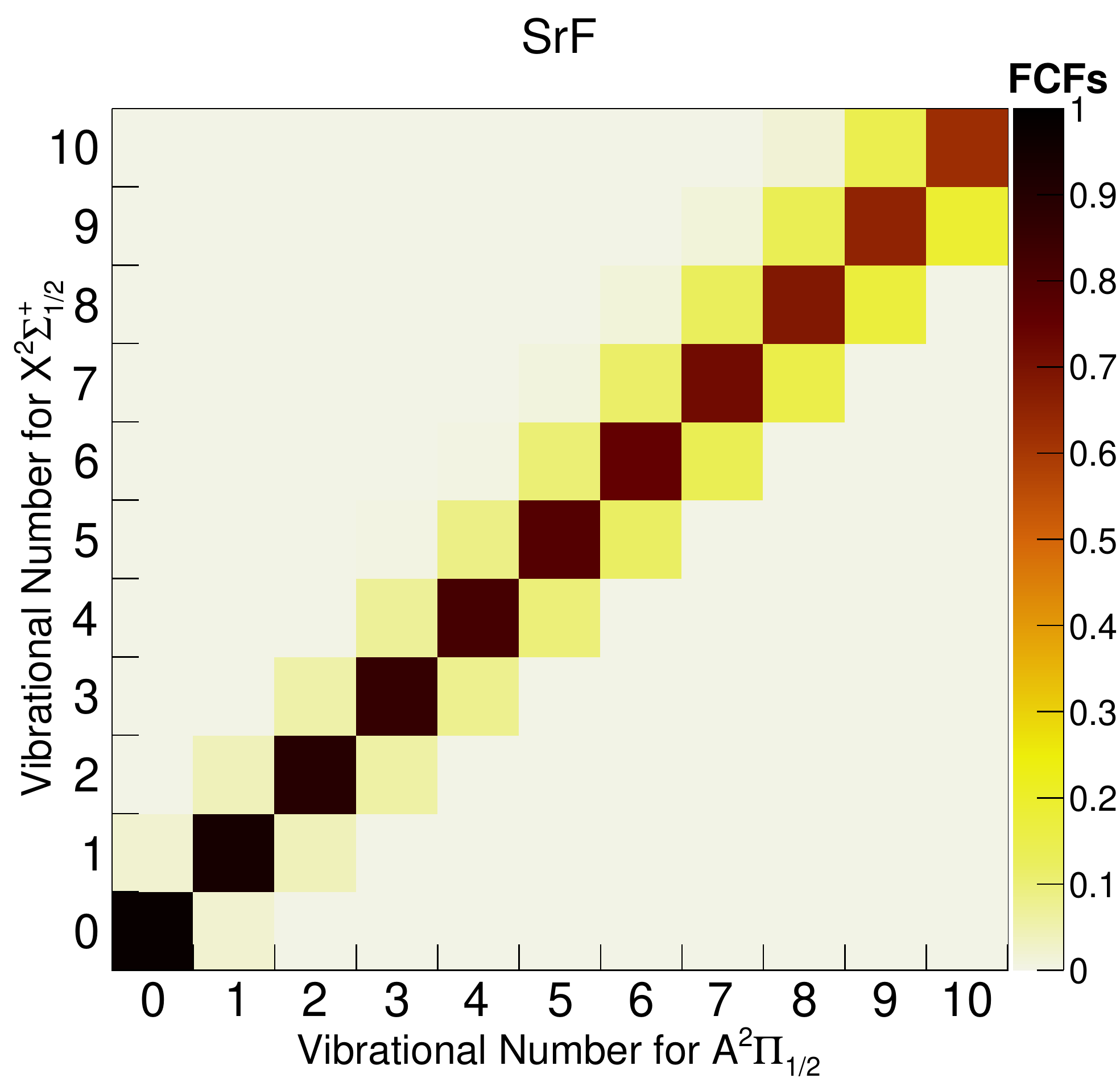}
\includegraphics[scale=0.75,width=0.32\linewidth]{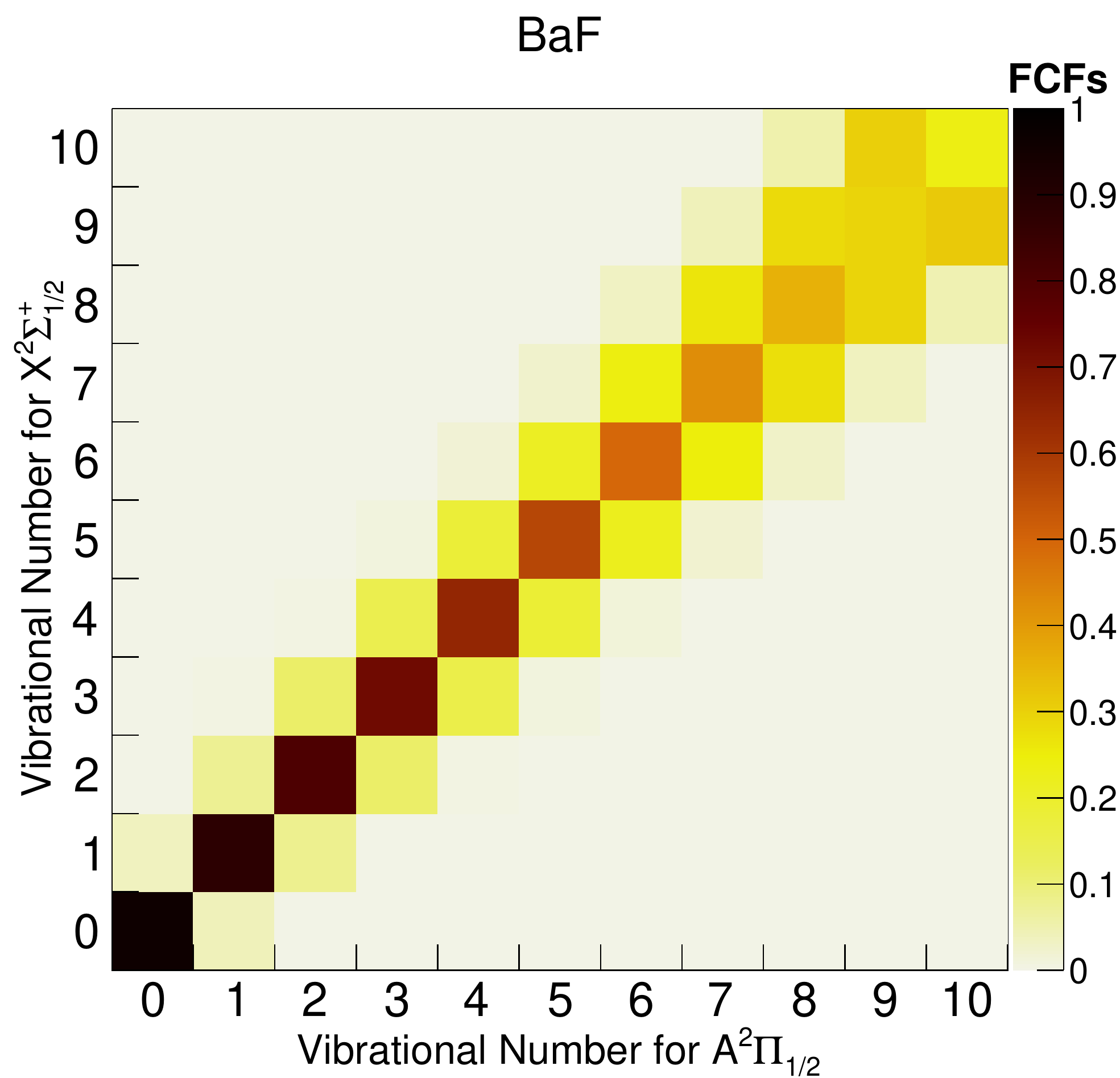}\\
\justifying 
\caption{Calculated Franck-Condon factors for the vibronic transitions between the $\vert A^2\Pi_{\frac{1}{2}},v'\rangle$ and the $\vert X^2\Sigma^{+}_{\frac{1}{2}},v\rangle$ states of CaF, SrF, and BaF (Color online).}
\label{FCFAX}
\end{figure*}

For the three molecules, the FCFs of the $A^2\Pi_{1/2}-X ^2\Sigma^{+}_{1/2}$ transitions (the intended cooling transition for BaF) exhibit a highly diagonal behaviour,  as can also be seen from Fig.~\ref{FCFAX}. This is due to the very similar equilibrium bond lengths of the ground and the $A^2{\Pi}$ states in all the molecules investigated here, and it makes these molecules excellent species for laser-cooling.

Wall \textit{et al.} \cite{Wall2008} have measured the FCF of the $A-X(0-0)$ band in CaF using the saturation of laser-induced fluorescence. Our result ($0.974$) is consistent with the experimental value ($0.968$-$1.000$).  We find that the diagonal FCF is largest for the SrF molecule, and the off-diagonal decay in the $(0-1)$ band the smallest. Our results for the diagonal $(0-0)$ and off-diagonal FCF for BaF ($0.960$ and $0.039$ respectively) predict a slightly less diagonal character for this system. Our calculations are also in good agreement with previous theoretical works \cite{Pelegrini2005,Barry2013,Karthikeyan2013,Chen2016s,Kang2016,Xu2017}.

\begin{table*}[t]
\caption{\label{BaF-FCF-A12} Frank-Condon factors (FCFs) for the vibronic transitions between the $\vert A^2\Pi_{\frac{1}{2}},v'\rangle$ and the $\vert X^2\Sigma^{+}_{\frac{1}{2}},v\rangle$ states of CaF, SrF, and BaF.}
\begin{subtable}{1\textwidth}
\sisetup{table-format=-1.2}
\begin{ruledtabular}
\begin{tabular}{c|ccccll}
\diagbox{$A^2{\Pi}_{\frac{1}{2}}$}{$X^2\Sigma^{+}_{\frac{1}{2}}$}&$v=0$&$v=1$&$v=2$&$v=3$&Method&Reference\\
\hline
${v}^{\prime}=0$&$9.739\times10^{-1}$&$2.523\times10^{-2}$&$8.742\times10^{-4}$&$3.588\times10^{-5}$&X2C-FSCC&This work\\
&$0.964$&$0.036$&$0.000$&$0.000$&MRCI&\cite{Pelegrini2005}\\
&$0.968-1.000$&&&&Experiment&\cite{Wall2008}\\
${v}^{\prime}=1$&$2.610\times10^{-2}$&$9.236\times10^{-1}$&$4.770\times10^{-2}$&$2.482\times10^{-3}$&X2C-FSCC&This work\\
&$0.035$&$0.895$&$0.070$&$0.000$&MRCI&\cite{Pelegrini2005}\\
${v}^{\prime}=2$&$4.427\times10^{-5}$&$5.107\times10^{-2}$&$8.763\times10^{-1}$&$6.760\times10^{-2}$&X2C-FSCC&This work\\
&$0.001$&$0.065$&$0.830$&$0.103$&MRCI&\cite{Pelegrini2005}\\
${v}^{\prime}=3$&$2.016\times10^{-7}$&$1.253\times10^{-4}$&$7.494\times10^{-2}$&$8.318\times10^{-1}$&X2C-FSCC&This work\\
&$0.000$&$0.004$&$0.092$&$0.767$&MRCI&\cite{Pelegrini2005}\\
\end{tabular}
\end{ruledtabular}
\caption{CaF}\label{tab:FCF1}
\end{subtable}
\bigskip
\begin{subtable}{1\textwidth}
\sisetup{table-format=-1.2} 
\begin{ruledtabular}
\begin{tabular}{c|ccccll}
\diagbox{$A^2{\Pi}_{\frac{1}{2}}$}{$X^2\Sigma^{+}_{\frac{1}{2}}$} & $v=0$ & $v=1$ & $v=2$ & $v=3$ &Method&Reference \\
\hline

${v}^{\prime}=0$ & $9.789\times10^{-1}$ & $2.054\times10^{-2}$ & $5.117\times10^{-4}$ & $1.530\times10^{-5}$& X2C-FSCC&This work\\
&$0.98$&0.018&$4.30\times10^{-4}$& $1.26\times10^{-5}$&RKR&\cite{Shuman2010,Barry2013}\\

${v}^{\prime}=1$ & $2.102\times10^{-2}$ & $9.377\times10^{-1}$ & $3.969\times10^{-2}$ & $1.489\times10^{-3}$& X2C-FSCC&This work\\
&0.019&0.945&0.035&0.001&RKR&\cite{Shuman2010,Barry2013}\\

${v}^{\prime}=2$ & $4.741\times10^{-5}$ & $4.158\times10^{-2}$ & $8.978\times10^{-1}$ & $5.749\times10^{-2}$& X2C-FSCC&This work\\
&$2.72\times10^{-5}$&0.037&$0.910$&0.051&RKR&\cite{Shuman2010,Barry2013}\\

${v}^{\prime}=3$ & $6.203\times10^{-9}$ & $1.411\times10^{-4}$ & $6.168\times10^{-2}$ & $8.592\times10^{-1}$& X2C-FSCC&This work\\
& $1.60\times10^{-8}$&$8.15\times10^{-5}$&0.054&$0.876$&RKR&\cite{Shuman2010,Barry2013}\\    
\end{tabular}
\end{ruledtabular}
\caption{SrF}\label{tab:FCF2}
\end{subtable}
\bigskip
\begin{subtable}{1\textwidth}
\sisetup{table-format=-1.2}
\begin{ruledtabular}
\begin{tabular}{c|ccccll}
\diagbox{$A^2{\Pi}_{\frac{1}{2}}$}{$X^2\Sigma^{+}_{\frac{1}{2}}$}&$v=0$&$v=1$&$v=2$&$v=3$&Method&Reference\\
\hline

${v}^{\prime}=0$ & $9.601\times10^{-1}$ & $3.892\times10^{-2}$ & $9.899\times10^{-4}$ & $1.318\times10^{-5}$& X2C-FSCC&This work\\
&$0.93$&$0.07$&&&RKR&\cite{Berg1993}\\
&$0.951$&$0.048$&$0.002$&$0.000$&MPM&\cite{Karthikeyan2013}\\
&$0.951$&$0.048$&0.002&$2.7\times10^{-5}$&RKR&\cite{Chen2016s}\\

&0.981&0.019&$3.96\times10^{-4}$& $2.98\times10^{-6}$&CASSCF+MRCI+SOC&\cite{Kang2016}\\
&$0.947$&$0.051$&$0.002$& $0.000$&MPM&\cite{Xu2017}\\

${v}^{\prime}=1$ & $3.923\times10^{-2}$ & $8.807\times10^{-1}$ & $7.695\times10^{-2}$ & $3.051\times10^{-3}$& X2C-FSCC&This work\\
&$0.049$&$0.854$&$0.093$&$0.005$&MPM&\cite{Karthikeyan2013}\\
&$0.048$&$0.854$&$0.093$&0.005&RKR&\cite{Chen2016s}\\

&0.019&0.940&0.039&$0.001$&CASSCF+MRCI+SOC&\cite{Kang2016}\\
&$0.052$&$0.845$&&&MPM&\cite{Xu2017}\\

${v}^{\prime}=2$ & $6.894\times10^{-4}$ & $7.812\times10^{-2}$ & $8.011\times10^{-1}$ & $1.137\times10^{-1}$& X2C-FSCC&This work\\
&$0.000$&$0.096$&$0.758$&$0.135$&MPM&\cite{Karthikeyan2013}\\
&$9.1\times10^{-4}$&$0.096$&$0.758$&$0.135$&RKR&\cite{Chen2016s}\\

&$7.10\times10^{-5}$&0.040&0.896&0.060&CASSCF+MRCI+SOC&\cite{Kang2016}\\

${v}^{\prime}=3$ & $3.405\times10^{-6}$ & $2.224\times10^{-3}$ & $1.162\times10^{-1}$ & $7.219\times10^{-1}$& X2C-FSCC&This work\\
&&$0.003$& $0.141$&$0.666$&MPM&\cite{Karthikeyan2013}\\
&$1.9\times10^{-6}$&$0.003$&$0.141$&$0.664$&RKR&\cite{Chen2016s}\\

&$1.14\times10^{-6}$&$2.88\times10^{-4}$&0.063&0.849&CASSCF+MRCI+SOC&\cite{Kang2016}\\
\end{tabular}
\end{ruledtabular}
  \caption{BaF}\label{tab:FCF3}
\end{subtable}
\end{table*}

\begin{table*}[t]
\caption{\label{BaF-FCF2} Frank-Condon factors (FCFs) using X2C-FSCC method for the vibronic transitions between the $\vert A^2\Pi_{\frac{1}{2}},v'\rangle$ and the $\vert A'^2\Delta_{\frac{3}{2}},v\rangle$ states of BaF (present work, X2C-FSCC).}
\begin{subtable}{1\textwidth}
\sisetup{table-format=-1.2} 
\begin{ruledtabular}
\begin{tabular}{l|cccc}
\diagbox{$A^2\Pi_{\frac{1}{2}}$}{$A'^2\Delta_{\frac{3}{2}}$} & $v=0$ & $v=1$ & $v=2$ & $v=3$ \\
\hline

${v}^{\prime}=0$ & $9.856\times10^{-1}$ & $1.404\times10^{-2}$ & $3.359\times10^{-4}$ & $9.470\times10^{-6}$\\

${v}^{\prime}=1$ & $1.438\times10^{-2}$ & $9.578\times10^{-1}$ & $2.685\times10^{-2}$ & $9.624\times10^{-4}$\\

${v}^{\prime}=2$ & $2.512\times10^{-6}$ & $2.818\times10^{-2}$ & $9.314\times10^{-1}$ & $3.851\times10^{-2}$\\

${v}^{\prime}=3$ & $1.443\times10^{-7}$ & $6.325\times10^{-6}$ & $4.143\times10^{-2}$ & $9.063\times10^{-1}$\\
\end{tabular}
\end{ruledtabular}

\end{subtable}
\end{table*}

\begin{table*}[t]
\caption{\label{BaF-FCF-ApX} Frank-Condon factors (FCFs) for the vibronic transitions between the $\vert A'^2\Delta_{\frac{3}{2}},v'\rangle$ and the $\vert X^2\Sigma^{+}_{\frac{1}{2}},v\rangle$  states of CaF, SrF, and BaF (present work, X2C-FSCC).\\}
\begin{subtable}{1\textwidth}
\sisetup{table-format=-1.2} 
\begin{ruledtabular}
\begin{tabular}{c|cccc}
\diagbox{$A'^2\Delta_{\frac{3}{2}}$}{$X^2\Sigma^{+}_{\frac{1}{2}}$} & $v=0$ & $v=1$ & $v=2$ & $v=3$ \\
\hline
${v}^{\prime}=0$ & $8.544\times10^{-1}$ & $1.354\times10^{-1}$ & $9.884\times10^{-3}$ & $4.105\times10^{-4}$\\         
${v}^{\prime}=1$ & $1.331\times10^{-1}$ & $6.018\times10^{-1}$ & $2.355\times10^{-1}$ & $2.801\times10^{-2}$\\
${v}^{\prime}=2$ & $1.180\times10^{-2}$ & $2.270\times10^{-1}$ & $4.009\times10^{-1}$ & $3.035\times10^{-1}$\\
${v}^{\prime}=3$ & $7.600\times10^{-4}$ & $3.271\times10^{-2}$ & $2.860\times10^{-1}$ & $2.475\times10^{-1}$\\
\end{tabular}
\end{ruledtabular}
\caption{CaF}\label{tab:FCF32}
\end{subtable}
\bigskip
\begin{subtable}{1\textwidth}
\sisetup{table-format=-1.2} 
\begin{ruledtabular}
\begin{tabular}{c|cccc}
\diagbox{$A'^2\Delta_{\frac{3}{2}}$}{$X^2\Sigma^{+}_{\frac{1}{2}}$} & $v=0$ & $v=1$ & $v=2$ & $v=3$\\
\hline

${v}^{\prime}=0$ & $9.696\times10^{-1}$ & $2.990\times10^{-2}$ & $4.924\times10^{-4}$ & $3.376\times10^{-6}$\\

${v}^{\prime}=1$ & $2.992\times10^{-2}$ & $9.089\times10^{-1}$ & $5.961\times10^{-2}$ & $1.522\times10^{-3}$\\

${v}^{\prime}=2$ & $4.660\times10^{-4}$ & $5.969\times10^{-2}$ & $8.477\times10^{-1}$ & $8.896\times10^{-2}$\\

${v}^{\prime}=3$ & $2.975\times10^{-6}$ & $1.476\times10^{-3}$ & $8.903\times10^{-2}$ & $7.863\times10^{-1}$\\
\end{tabular}
\end{ruledtabular}
\caption{SrF}\label{tab:FCF22}
\end{subtable}
\bigskip
\begin{subtable}{1\textwidth}
\sisetup{table-format=-1.2} 
\begin{ruledtabular}
\begin{tabular}{c|cccc}
\diagbox{$A'^2\Delta_{\frac{3}{2}}$}{$X^2\Sigma^{+}_{\frac{1}{2}}$} & $v=0$ & $v=1$ & $v=2$ & $v=3$ \\
\hline

${v}^{\prime}=0$ & $9.007\times10^{-1}$ & $9.480\times10^{-2}$ & $4.361\times10^{-3}$ & $1.087\times10^{-4}$\\

${v}^{\prime}=1$ & $9.333\times10^{-2}$ & $7.196\times10^{-1}$ & $1.739\times10^{-1}$ & $1.269\times10^{-2}$\\

${v}^{\prime}=2$ & $5.678\times10^{-3}$ & $1.683\times10^{-1}$ & $5.623\times10^{-1}$ & $2.381\times10^{-1}$\\

${v}^{\prime}=3$ & $2.597\times10^{-4}$ & $1.619\times10^{-2}$ & $2.261\times10^{-1}$ & $4.276\times10^{-1}$\\
\end{tabular}
\end{ruledtabular}
\caption{BaF}\label{tab:FCF12}
\end{subtable}
\end{table*}

\begin{figure}[t]
\centering

\includegraphics[scale=0.95,width=0.95\linewidth]{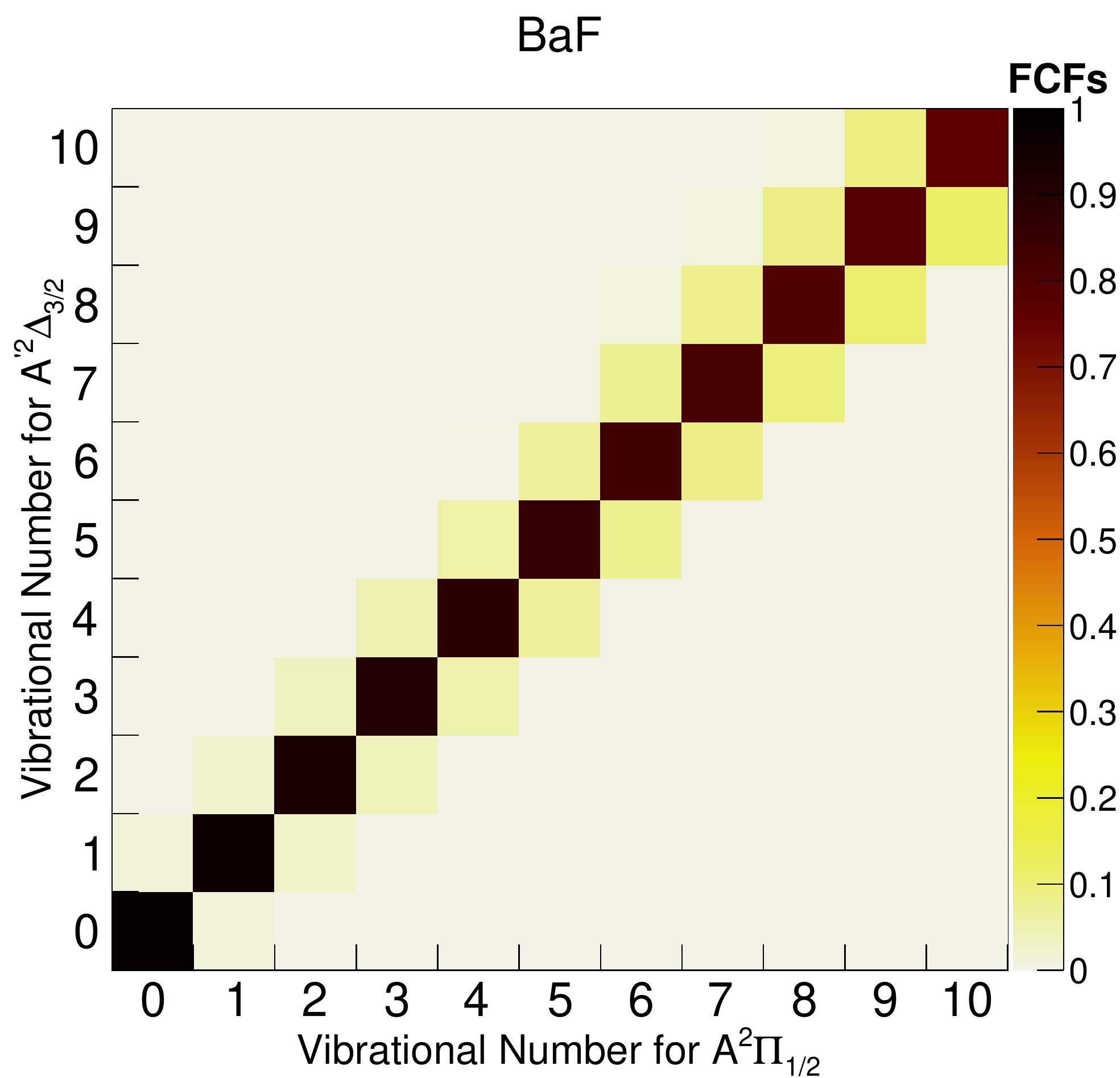}\\
\justifying 
\caption{Calculated Franck-Condon factors for the vibronic transitions between the $\vert A'^2\Delta_{\frac{3}{2}},v\rangle$ and the $\vert A^2\Pi_{\frac{1}{2}},v'\rangle$ states of BaF (Color online).}\label{Fig_D32_P12}
\end{figure}

The $A-A'$ transition constitutes a possible leak in the cooling cycle of BaF; for CaF and SrF the $A' \Delta_{3/2}$ state is higher than the $A ^2\Pi_{1/2}$ and therefore not a concern in this context. To the best of our knowledge no previous calculations or measurements were performed for the FCFs between these two states. The FCF of the $A-A'(0-0)$ transition in BaF is 0.986 (Table \ref{BaF-FCF2} and Fig.~\ref{Fig_D32_P12}), due to the similar equilibrium bond length of the two states. We also present the FCFs for the decay of the $A'^2{\Delta}$ states of the three species to the ground state (Table ~\ref{BaF-FCF-ApX}).  Implications of these results for the laser-cooling of BaF are discussed below.

\begin{table*}[t]
\caption{\label{BaF-FCF3}  Frank-Condon factors (FCFs) for the vibronic transitions between the $\vert B^2\Sigma^{+}_{\frac{1}{2}},v'\rangle$ and the $\vert X^2\Sigma^{+}_{\frac{1}{2}},v\rangle$ states of CaF, SrF, and BaF (present work, X2C-FSCC).}
\begin{subtable}{1\textwidth}
\sisetup{table-format=-1.2}
\begin{ruledtabular}
\begin{tabular}{c|cccc}
\diagbox{$B^2\Sigma^{+}_{\frac{1}{2}}$}{$X^2\Sigma^{+}_{\frac{1}{2}}$} & $v=0$ & $v=1$ & $v=2$ & $v=3$ \\
\hline
${v}^{\prime}=0$ & $9.992\times10^{-1}$ & $7.270\times10^{-4}$ & $3.809\times10^{-5}$ & $9.834\times10^{-8}$\\        
${v}^{\prime}=1$ & $7.396\times10^{-4}$ & $9.973\times10^{-1}$ & $1.814\times10^{-3}$ & $1.176\times10^{-4}$\\
${v}^{\prime}=2$ & $2.473\times10^{-5}$ & $1.873\times10^{-3}$ & $9.945\times10^{-1}$ & $3.322\times10^{-3}$\\
${v}^{\prime}=3$ & $7.981\times10^{-7}$ & $6.775\times10^{-5}$ & $3.481\times10^{-3}$ & $9.907\times10^{-1}$\\
\end{tabular}
\end{ruledtabular}
\caption{CaF}\label{tab:FCF1}
\end{subtable}
\bigskip
\begin{subtable}{1\textwidth}
\sisetup{table-format=-1.2}
\begin{ruledtabular}
\begin{tabular}{c|cccc}
\diagbox{$B^2\Sigma^{+}_{\frac{1}{2}}$}{$X^2\Sigma^{+}_{\frac{1}{2}}$} & $v=0$ & $v=1$ & $v=2$ & $v=3$ \\
\hline

${v}^{\prime}=0$ & $9.961\times10^{-1}$ & $3.866\times10^{-3}$ & $3.604\times10^{-6}$ & $7.685\times10^{-9}$\\

${v}^{\prime}=1$ & $3.856\times10^{-3}$ & $9.881\times10^{-1}$ & $8.000\times10^{-3}$ & $1.190\times10^{-5}$\\

${v}^{\prime}=2$ & $1.343\times10^{-5}$ & $7.959\times10^{-3}$ & $9.796\times10^{-1}$ & $1.241\times10^{-2}$\\

${v}^{\prime}=3$ & $7.913\times10^{-8}$ & $4.258\times10^{-5}$ & $1.231\times10^{-2}$ & $9.705\times10^{-1}$\\
\end{tabular}
\end{ruledtabular}
\caption{SrF}\label{tab:FCF2}
\end{subtable}
\bigskip
\begin{subtable}{1\textwidth}
\sisetup{table-format=-1.2} 
\begin{ruledtabular}
\begin{tabular}{c|cccc}
\diagbox{$B^2\Sigma^{+}_{\frac{1}{2}}$}{$X^2\Sigma^{+}_{\frac{1}{2}}$}&$v=0$&$v=1$&$v=2$&$v=3$\\
\hline

${v}^{\prime}=0$ & $7.995\times10^{-1}$ & $1.811\times10^{-1}$ & $1.832\times10^{-2}$ & $1.073\times10^{-3}$\\
&$0.81^a$ & $0.17^a$&&\\

${v}^{\prime}=1$ & $1.760\times10^{-1}$ & $4.782\times10^{-1}$ & $2.924\times10^{-1}$ & $4.915\times10^{-2}$\\

${v}^{\prime}=2$ & $2.221\times10^{-2}$ & $2.751\times10^{-1}$ & $2.570\times10^{-1}$ & $3.482\times10^{-1}$\\

${v}^{\prime}=3$ & $2.105\times10^{-3}$ & $5.717\times10^{-2}$ & $3.156\times10^{-1}$ & $1.159\times10^{-1}$\\
\end{tabular}
\end{ruledtabular}
 \caption{BaF}\label{tab:FCF3}
   \begin{tablenotes}
      \footnotesize
      \item[\emph{a}]{$^a$ Previous study using RKR method \cite{Berg1993}.}
  \end{tablenotes}
\end{subtable}
\end{table*}

\begin{figure*}[t]
\centering
\includegraphics[scale=0.75,width=0.32\linewidth]{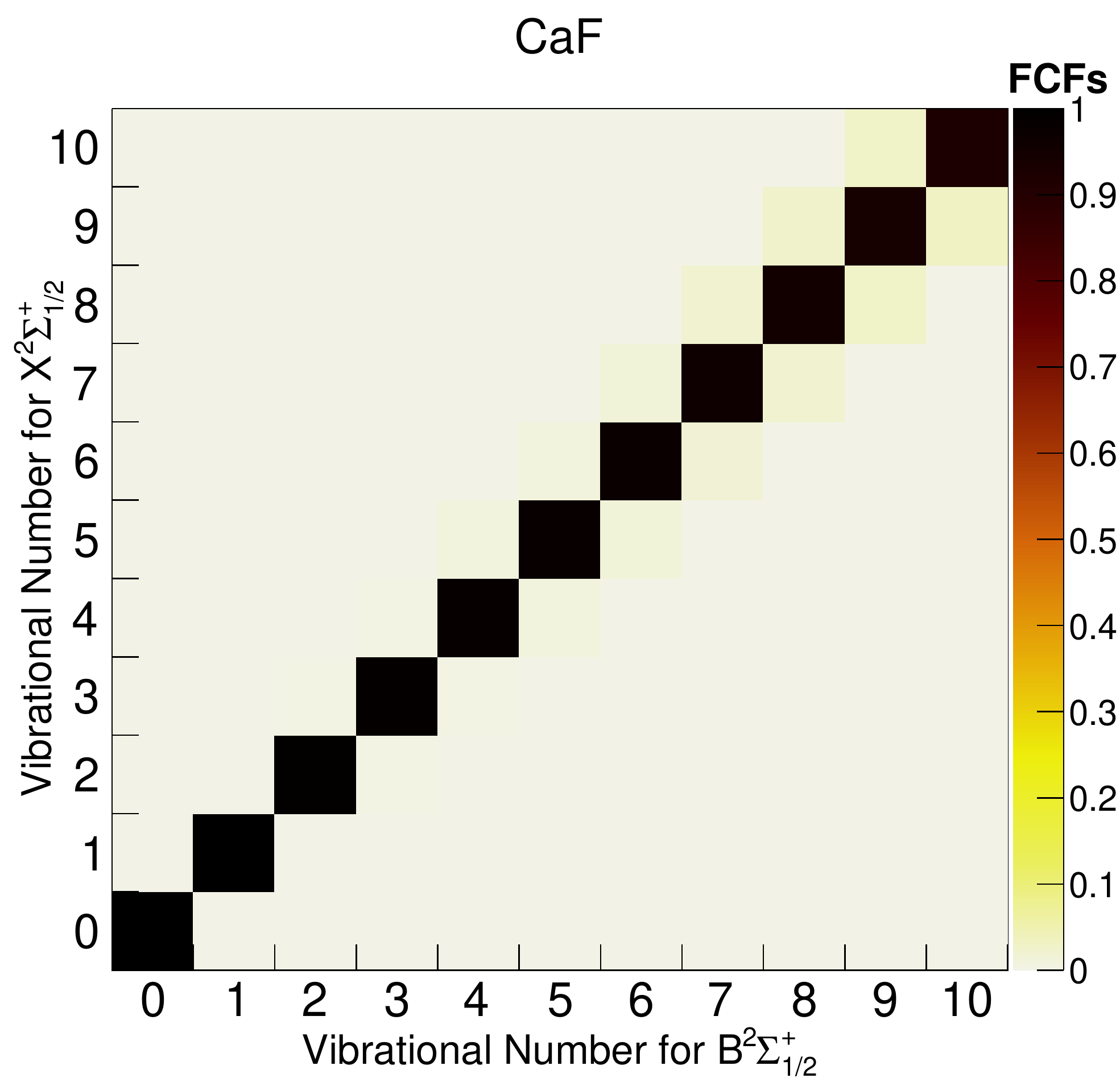}
\includegraphics[scale=0.75,width=0.32\linewidth]{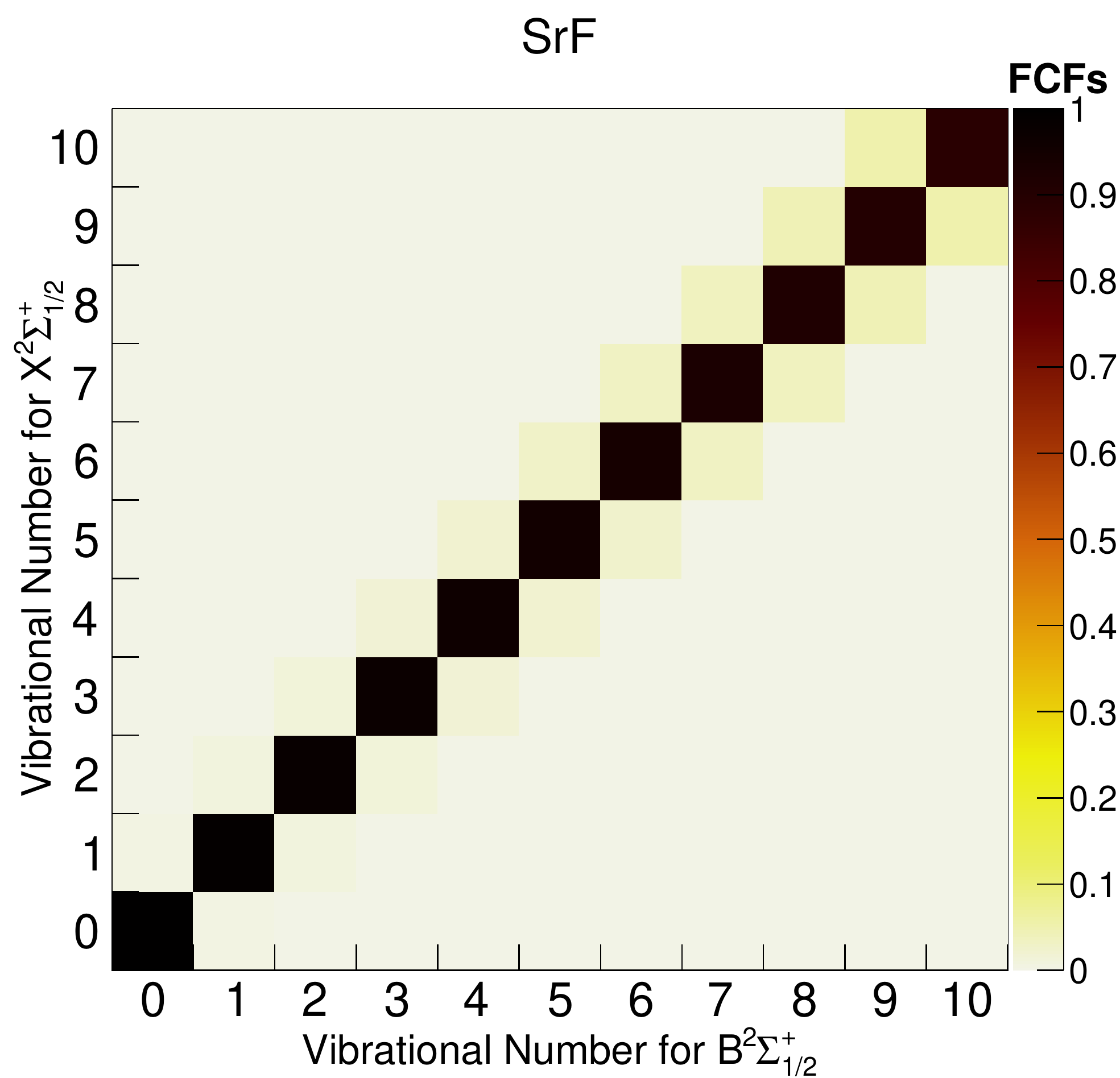}
\includegraphics[scale=0.75,width=0.32\linewidth]{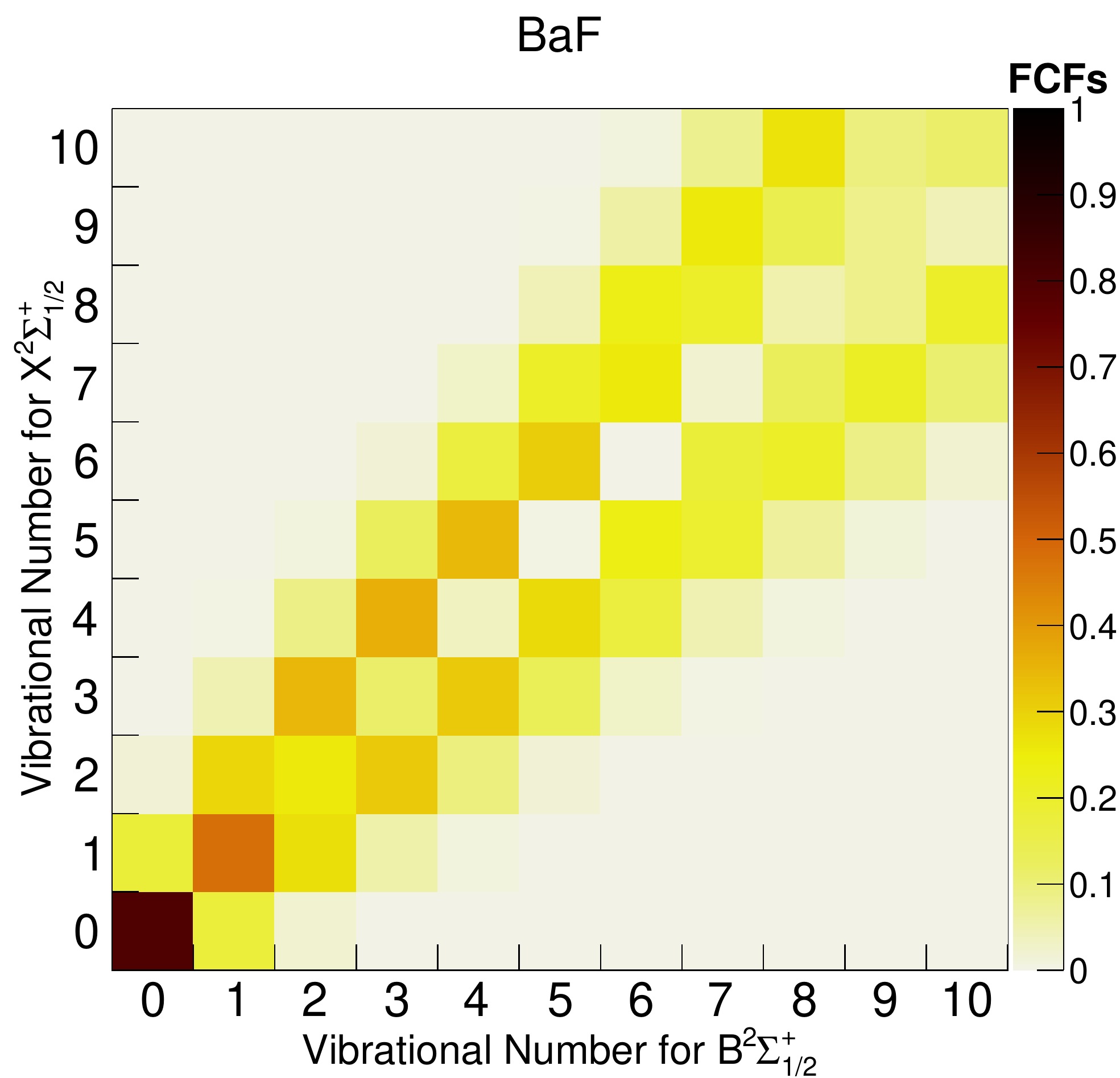}\\
\justifying
\caption{Calculated Franck-Condon factors for the vibronic transitions between the $\vert B^2\Sigma^{+}_{\frac{1}{2}},v'\rangle$ and the $\vert X^2\Sigma^{+}_{\frac{1}{2}},v\rangle$ states of CaF, SrF, and BaF (Color online).}
\label{FCFBX}
\end{figure*}

The $B ^2\Sigma^{+}_{1/2} - X ^2 \Sigma^{+}_{1/2}$ transition was demonstrated as an alternative cooling route for CaF \cite{Truppe2017b}. We thus explore the FCFs of this transition in the three molecules (Table~\ref{BaF-FCF3} and Fig.~\ref{FCFBX}). In case of CaF, the FCFs are indeed highly diagonal, with the $B-X(0-0)$ FCF extremely close to unity, and in SrF it is 0.996. BaF, however, has an FCF of about 0.800 for the same transition, caused by a significantly larger $R_e$ of the $B ^2 \Sigma^{+}_{1/2}$ state compared to the ground state.   

\subsection{Static and Transition Dipole Moments}

The calculated DMs at experimental bond lengths $R_e$ are given in Tables ~\ref{PDMsCa}, ~\ref{PDMsSr}, and ~\ref{PDMsBa}  and compared to experimental values (where available) and to previous theoretical investigations.

\begin{table}[t]
\caption{Calculated dipole moments (a.u.) of CaF at the experimental bond length $R_e$, compared to previous calculations and experiment.\label{PDMsCa}}
\begin{ruledtabular}
\begin{tabular}{llll}
State&DM&Method&Reference\\\hline
$X^2\Sigma^{+}_{1/2}$ &1.18&X2C-MRCISD&This work\\
                    &1.25&X2C-FSCC&This work\\
                    &1.31&Ionic model&\cite{Torring1984}\\
                    &1.18&LFM&\cite{Rice1985}\\
                    &1.02&CISD&\cite{Langhoff1986}\\
                    &1.26&EPM&\cite{Mestdagh1991}\\
                    &1.18&MRCI&\cite{Bundgen1991}\\
                    &1.32&LFM&\cite{Allouche1993}\\
                    &1.26&MP2&\cite{Buckingham1993}\\
                    &1.04&FDHF&\cite{Kobus2000}\\                 
                    &1.24&EOVERM&\cite{Raouafi2001}\\
                    &1.20&RCCSD(T)&\cite{Harrison2002}\\
                    &1.24&RCCSD&\cite{Prasannaa2016}\\
                    &1.30$^a$&CASSCF+MRCI&\cite{El2017}\\
                    &1.26&RCCSD(T)&\cite{Abe2018}\\
                    &1.20&RCCSD(T)&\cite{Hou2018}\\
                    
                    &1.21(3)&Experiment&\cite{Childs1984}\\

$A'^2\Delta_{3/2}$  &2.44&X2C-MRCISD&This work\\
                    &2.57&X2C-FSCC&This work\\
$A'^2\Delta_{5/2}$  &2.44&X2C-MRCISD&This work\\
                    &2.57&X2C-FSCC&This work\\ 
$A'^2\Delta$        &2.98&LFM&\cite{Rice1985}\\ 
                    &3.04&EPM&\cite{Torring1989}\\
                    &3.2&CASSCF+MRCI&\cite{El2017}\\

$A^2\Pi_{1/2}$ 	    &1.04&X2C-MRCISD&This work\\
                    &1.08&X2C-FSCC&This work\\
$A^2\Pi_{3/2}$ 	    &1.04&X2C-MRCISD&This work\\
                    &1.08&X2C-FSCC&This work\\
$A^2\Pi$            &1.61&LFM&\cite{Rice1985}\\ 
                    &1.01&EPM&\cite{Torring1989}\\
                    &1.00&EOVERM&\cite{Raouafi2001}\\ 
                    &0.96(2)&Experiment&\cite{Ernst1989}\\ 

$B^2\Sigma^{+}_{1/2}$&0.69&X2C-MRCISD&This work\\
                    &0.89&X2C-FSCC&This work\\
                    &2.25&LFM&\cite{Rice1985}\\ 
                    &0.63&EPM&\cite{Torring1989}\\
                    &0.73&EOVERM&\cite{Raouafi2001}\\
\end{tabular}
\end{ruledtabular}
\begin{tablenotes}
\footnotesize
\item[]{$^a$ This is evaluated around the equilibrium bond distance from Fig. 4 of Ref. \cite{El2017}.}
\end{tablenotes}
\end{table}

The majority of previous theoretical investigations of the DMs of these molecules were carried out in a nonrelativistic framework; the only exception being the relativistic coupled cluster studies of the ground state DMs of the three molecules \cite{Prasannaa2016,Abe2018,Hou2018,Harrison2002,Sasmal2015,Fazil2018}. This is the first relativistic study of the DMs of the excited states. We have performed the calculation using two approaches: FSCC and MRCI. The results obtained using the two methods are within a few percent of each other for most of the states considered here, with the exception of the $B ^2\Sigma^{+}_{1/2}$ states of the three molecules, where the differences are significantly larger.

In case of the ground state DMs our results are generally in good agreement with the majority of the earlier theoretical publications (in particular, as expected, with the relativistic coupled cluster values \cite{Harrison2002,Prasannaa2016,Abe2018,Hou2018,Sasmal2015,Fazil2018}), and within 10\% of the measured values. For the excited states, previous data are more scarce.
For the $A^2\Pi$ state in CaF and SrF, our FSCC and MRCI results overestimate the experimental values somewhat (5-12\%); the error is smaller for MRCI. In case of the $B^2 \Sigma^{+}_{1/2}$ state of SrF, FSCC performs on the same level, but the MRCI results are too low almost by a factor of 2; this is consistent with the deviation of the MRCI and the FSCC values for this state in all the molecules. For BaF, there is no experiment available for the DMs of the excited states. For these states, our DMs are generally lower than those from the earlier calculations, with the best agreement obtained where MRCI approach was also employed \cite{Tohme2015,Kang2016}; the discrepancy can be attributed to neglect of relativistic effects in the previous works, or the use of a significantly smaller basis set in Ref. \cite{Tohme2015}. We expect the present predictions to be the most accurate, due to the quality of the methods employed here.

The good agreement of the MRCI DM results with the FSCC values (which are expected to be more accurate) and with experiment validates the use of this method for calculation of the TDMs, where FSCC is not yet applicable.

The calculated transition dipole moments between the ground and the excited states and in between the different excited states are collected in Table ~\ref{TDMs}. Experimental verification of the TDM values can be obtained from comparison with measured lifetimes of excited states, as discussed in the next subsection. Also, good agreement is found with previous theoretical investigations, in particular where MRCI approach was used \cite{Tohme2015,Kang2016}. The new results presented here are, however, the first relativistic calculations of the TDMs of these molecules.

No experimental and limited theoretical information is available for the TDMs between the $A ^2 \Pi_{1/2}$ and the $A'^2 \Delta_{3/2}$ states, the latter being a possible leak channel in the laser-cooling cycle. In case of CaF, our prediction is somewhat higher than the ligand field method calculation of Ref. \cite{Rice1985}, but of particular note is the discrepancy of almost two orders of magnitude with the predictions of Kang \textit{et al.} \cite{Kang2016} for BaF. Our predicted TDM for this transition is 2.33 a.u., which is close to that of CaF and SrF, as expected. The low value presented in Ref. \cite{Kang2016} (0.04 a.u.) is appropriate for a forbidden transition, which is not the case for $A ^2 \Pi_{1/2}$ to $A'^2 \Delta_{3/2}$; thus, we view the present prediction as more reliable. We note that the avoided crossing between the $A ^2 \Pi_{3/2}$ and $A'^2 \Delta_{3/2}$ states of BaF, an artifact introduced by the MRCI method, somewhat lowers the expected accuracy of the TDMs for the weak transitions $A'^2 \Delta_{5/2} - A'^2 \Delta_{3/2}$, $X^2 \Sigma_{1/2} - A'^2 \Delta_{3/2}$, $A^2 \Pi_{3/2} - A'^2 \Delta_{3/2}$ and $A^2 \Pi_{3/2} - A^2 \Pi_{1/2}$. Comparing results obtained from different basis sets, we estimate the size of this error to be up to 20\%. This only affects the $A'^2 \Delta_{5/2}$ and $A'^2 \Delta_{3/2}$ lifetimes of BaF presented in the following subsection.

\begin{table}[b]
\caption{Calculated dipole moments (a.u.) of SrF at the experimental bond length $R_e$, compared to previous calculations and experiment.\label{PDMsSr}}
\begin{ruledtabular}
\begin{tabular}{llll}
State&DM&Method&Reference\\\hline
$X^2\Sigma^{+}_{1/2}$&1.26&X2C-MRCISD&This work\\
                    &1.36&X2C-FSCC&This work\\
                    &1.44&Ionic model&\cite{Torring1984}\\ 
                    &0.99&CISD&\cite{Langhoff1986}\\
                    &1.42&EPM&\cite{Mestdagh1991}\\
                    &1.49&LFM&\cite{Allouche1993}\\
                    &1.01&FDHF&\cite{Kobus2000}\\
                    &1.42& CASSCF+RSPT2 & \cite{Jardali2014}\\
                    &1.32& CASSCF+MRCI & \cite{Jardali2014}\\
                    &1.36&RCCSD&\cite{Sasmal2015}\\
                    &1.42&CCSD&\cite{Prasannaa2016}\\
                    &1.42&RCCSD(T)&\cite{Abe2018}\\
                    &1.38&RCCSD &\cite{Fazil2018} \\
                    &1.3643(4)&Experiment&\cite{Ernst1985}\\

$A'^2\Delta_{3/2}$ 	&2.39&X2C-MRCISD&This work\\
                    &2.50&X2C-FSCC&This work\\
$A'^2\Delta_{5/2}$ 	&2.39&X2C-MRCISD&This work\\
                    &2.50&X2C-FSCC&This work\\
$A'^2\Delta$        &3.36&EPM&\cite{Torring1989}\\
                    &3.18&LFM&\cite{Allouche1993}\\
                    &3.27&CASSCF+MRCI&\cite{Jardali2014}\\           

$A^2\Pi_{1/2}$ 		&0.85&X2C-MRCISD&This work\\
                    &0.91&X2C-FSCC&This work\\
$A^2\Pi_{3/2}$ 		&0.82&X2C-MRCISD&This work\\
                    &0.88&X2C-FSCC&This work\\
$A^2\Pi$            &0.85&EPM&\cite{Torring1989}\\
                    &1.29&LFM&\cite{Allouche1993}\\
                    &1.53&CASSCF+RSPT2&\cite{Jardali2014}\\
                    &1.64&CASSCF+MRCI&\cite{Jardali2014}\\
                    &0.81(2)&Experiment&\cite{Kandler1989}\\

$B^2\Sigma^{+}_{1/2}$&0.19&X2C-MRCISD&This work\\  
                    &0.40&X2C-FSCC&This work\\
                    &0.41&EPM&\cite{Torring1989}\\
                    &1.33&LFM&\cite{Allouche1993}\\
                    &1.26&CASSCF+MRCI&\cite{Jardali2014}\\
                    &0.36(2)&Experiment&\cite{Kandler1989}\\
\end{tabular}
\end{ruledtabular}
\end{table}

\begin{table}[t]
\caption{Calculated dipole moments (a.u.) of BaF at the experimental bond length $R_e$, compared to previous calculations and experiment.\label{PDMsBa}}
\begin{ruledtabular}
\begin{tabular}{llll}
State&DM&Method&Reference\\\hline
$X^2\Sigma^{+}_{1/2}$&1.14&X2C-MRCISD&This work\\
                    &1.27&X2C-FSCC&This work\\
                    &1.26&RASCI&\cite{Nayak2006}\\ 
                    &1.35&Ionic model&\cite{Torring1984}\\ 
                    &1.54&LFM&\cite{Allouche1993}\\
                    &1.15&AREP-RASSCF&\cite{Kozlov1997}\\
                    &1.16&CASSCF+MRCI&\cite{Tohme2015}\\
                    &1.33 &CASSCF+MRCI+SOC&\cite{Kang2016}\\
                    &1.34&RCCSD&\cite{Prasannaa2016}\\
                    &1.34&RCCSD(T)&\cite{Abe2018}\\  
                    &1.34&RCCSD &\cite{Fazil2018}\\
                    &1.247(1)&Experiment&\cite{Ernst1986}\\
                    
$A'^2\Delta_{3/2}$ 	&2.31&X2C-MRCISD&This work\\
                    &2.38&X2C-FSCC&This work\\
$A'^2\Delta_{5/2}$ 	&2.31&X2C-MRCISD&This work\\
                    &2.38&X2C-FSCC&This work\\
 $A'^2\Delta$       &3.57&EPM&\cite{Torring1989}\\
                    &3.31&LFM&\cite{Allouche1993}\\
 	                &2.47&CASSCF+MRCI&\cite{Tohme2015}\\
                    &2.64 &CASSCF+MRCI&\cite{Kang2016}\\
                    
$A^2\Pi_{1/2}$ 	    &0.40&X2C+MRCISD&This work\\
                    &0.53&X2C-FSCC&This work\\
$A^2\Pi_{3/2}$ 		&0.34&X2C+MRCISD&This work\\
                    &0.47&X2C-FSCC&This work\\
 $A^2\Pi$           &1.95&EPM&\cite{Torring1989}\\
                    &1.36&LFM&\cite{Allouche1993}\\
                    &0.86&CASSCF+MRCI&\cite{Tohme2015}\\
                    &1.01 &CASSCF+MRCI&\cite{Kang2016}\\
                    
$B^2\Sigma^{+}_{1/2}$&0.58&X2C-MRCISD&This work\\
                    &0.32&X2C-FSCC&This work\\
                    &1.61&EPM&\cite{Torring1989}\\
                    &1.31&LFM&\cite{Allouche1993}\\
                    &0.54&CASSCF+MRCI&\cite{Tohme2015}\\
\end{tabular}
\end{ruledtabular}
\end{table}

\begin{table}[t]
\caption{Calculated transition dipole moments (a.u.) between state 1 and state 2 at the ground state experimental bond length $R_e$. Present results in italics, experimental values in bold font.\label{TDMs}
}
\setlength{\tabcolsep}{3pt}
\begin{tabular}{llllll}
\hline\hline
\multirow{2}{*}{State 1}	&\multicolumn{5}{c}{State 2}\\
\cline{2-6}
&$A^2\Pi_{1/2}$	&$A^2\Pi_{3/2}$	&$A'^2\Delta_{3/2}$	&$A'^2\Delta_{5/2}$	&$B^2\Sigma^{+}_{1/2}$	\\
\hline
\textbf{CaF}\\
$X^2\Sigma^{+}_{1/2}$	&\textit{2.406}	&\textit{2.406}	&\textit{0.004}	&	&\textit{1.881}	\\
	&\multicolumn{2}{c}{2.32*$^a$}	&	&	&1.73$^a$	\\
	&\multicolumn{2}{c}{2.17*$^b$}	&	&	&1.64$^b$	\\
	&\multicolumn{2}{c}{2.34*$^c$}	&	&	&1.85$^c$	\\
	&	&	&	&	&1.79$^d$	\\
	&\multicolumn{2}{c}{\textbf{2.34}*$^e$}	&	&	&\textbf{1.71}$^e$	\\
$A^2\Pi_{1/2}$	&	&\textit{0.012}	&\textit{2.473}	&	&\textit{0.373}	\\
$A^2\Pi_{3/2}$	&	&	&\textit{0.000}	&\textit{2.476}	&\textit{0.370}	\\
$A^2\Pi$	&	&	&\multicolumn{2}{c}{1.76*$^a$}	&	\\
$A'^2\Delta_{3/2}$	&	&	&	&\textit{0.001}	&\textit{0.036}	\\
\hline
\textbf{SrF}\\
$X^2\Sigma^{+}_{1/2}$	&\textit{2.626}	&\textit{2.627}	&\textit{0.012}	&	&\textit{2.054}	\\
	&\multicolumn{2}{c}{2.37*$^b$}	&	&	&1.86$^b$\\
	&\multicolumn{2}{c}{\textbf{2.45}*$^e$}	&	&	&	\\
	&	&	&	&	&\textbf{2.45}$^f$	\\
$A^2\Pi_{1/2}$	&	&\textit{0.035}	&\textit{2.711}	&	&\textit{0.210}	\\
$A^2\Pi_{3/2}$	&	&	&\textit{0.004}	&\textit{2.728}	&\textit{0.195}	\\
$A'^2\Delta_{3/2}$	&	&	&	&\textit{0.009}	&\textit{0.173}	\\
\hline
\textbf{BaF}\\
$X^2\Sigma^{+}_{1/2}$	&\textit{2.810}	&\textit{2.797}	&\textit{0.272}	&	&\textit{2.226}	\\
	&\multicolumn{2}{c}{2.18*$^b$}	&	&	&1.85$^b$	\\
	&\multicolumn{2}{c}{3.20*$^\dagger$$^g$}	&	&	&2.40$^\dagger$$^g$	\\
	&\multicolumn{2}{c}{2.73*$^h$}	&\multicolumn{2}{c}{0.20*$^h$}	&	\\
	&	&\textbf{2.57}$^{i}$	&	&	&\textbf{2.10}$^i$		\\
	&\textbf{2.41}$^{j}$	&	&	&	&	\\
$A^2\Pi_{1/2}$	&	&\textit{0.242}	&\textit{2.332}	&	&\textit{0.100}	\\
$A^2\Pi_{3/2}$	&	&	&\textit{0.166}	&\textit{2.375}	&\textit{0.178}	\\
$A^2\Pi$	&	&	&\multicolumn{2}{c}{0.04*$^h$}	&	\\
$A'^2\Delta_{3/2}$	&	&	&	&\textit{0.193}	&\textit{0.316}	\\
\hline\hline
\end{tabular}
\begin{tablenotes}
\footnotesize
\item[]{*$\Omega$-unresolved transitions; $^\dagger$This is evaluated around the equilibrium bond distance from Fig. 10 of Ref. \cite{Tohme2015}; $^a$LFM \cite{Rice1985}; $^b$LFM \cite{Allouche1993}; $^c$EOVERM \cite{Raouafi2001}; $^d$MRCI \cite{El2017}; $^e$Experiment \cite{Dagdigian1974}; $^f$Experiment \cite{Berg1996}; $^g$CASSCF+MRCI \cite{Tohme2015}; $^h$CASSCF+MRCI \cite{Kang2016}; $^i$Experiment \cite{Berg1993}; $^j$Experiment \cite{Berg1998}.}
\end{tablenotes}
\end{table}

\subsection{Lifetimes of Excited States}

The transition rate of a vibronic transition is defined as 
\begin{equation}
\Gamma_{n'v'n''v''} = \frac{16\pi^3e^2a_{B}^{2}}{3h\epsilon_0}\nu^{3}_{n'v'n''v''}|\langle v'|M_{n'n''}(R)|v''\rangle|^2.
\label{rate}
\end{equation}
Here, $n'v'$ and $n''v''$ denote the upper and lower vibronic states (with $n$ for the electronic and $v$ for the vibrational part), $h$ is the Planck constant, $a_B$ is the Bohr radius, $\epsilon_0$ is the permittivity of free space, $M_{n'n''}(R)$ is the electronic TDM function, and $\nu_{n'v'n''v''}$ is the corresponding transition frequency. In the Franck-Condon (FC) approximation, one assumes the TDM to be independent of $R$, such that the integral can be factorized to become \cite{Larsson1983,Rui2014}
\begin{eqnarray}
\Gamma_{n'v'n''v''} \simeq & \frac{16\pi^3e^2a_{B}^{2}}{3h\epsilon_0}\nu^{3}_{n'v'n''v''}|\langle v'|v''\rangle|^2M_{n'n''}^2 \nonumber\\
= & \frac{16\pi^3e^2a_{B}^{2}}{3h\epsilon_0}\nu^{3}_{n'v'n''v''}q_{v'v''}M_{n'n''}^2.
\label{FCrate}
\end{eqnarray}
The squared overlaps of vibrational wavefunctions $|\langle v'|v''\rangle|^2 = q_{v'v''}$ are the FCFs obtained in Section \ref{sec:FCFs}. The transition rates $\Gamma_{n'v'n''v''}$ were calculated using the program LEVEL16 \cite{le2017level} and were subsequently used to calculate the lifetimes.

The lifetime $\tau_{n'v'}$ of an excited level can be derived by summing over all vibronic decay channels:
\begin{equation}
\tau_{n'v'} = \frac{1}{\sum\limits_{n''v''} \Gamma_{n'v'n''v''}}.
\label{lifetime}
\end{equation}
All lifetimes listed below were calculated from the transition rates according to Eq. \ref{rate}. However, the FC approximation would also be very appropriate for these molecules, as all errors that would be introduced in the transition rates by the FC approximation lie below 3.5\%. This includes the values for the branching ratios of relevance for laser-cooling. This justifies the use of FCFs for the interpretation of the investigated transitions in Sections \ref{sec:FCFs} and \ref{Lasercooling}.

The lifetimes of the excited states of CaF, SrF, and BaF are listed in Table~\ref{Lifetime}. The calculated lifetimes are lower by ~15-30 \% than the experimental values \cite{Dagdigian1974,Berg1993,Berg1996,Berg1998}, with the discrepancies highest for BaF (the uncertainty on the experimental CaF and SrF lifetimes was estimated as $\sim$2-4 ns \cite{Dagdigian1974}, and as low as  $\sim$1 ns for BaF \cite{Berg1993,Berg1998}). Furthermore, the calculated difference between the $A^2\Pi_{1/2}$ and the $A^2\Pi_{3/2}$ lifetimes is lower than that obtained in the experiment. Interestingly, for CaF the experimental lifetimes of the two states differ by 3.5 ns, which is higher than the corresponding difference in SrF (1.5 ns), in spite of CaF being a lighter system.  A new measurement of the lifetimes in question would thus be instrumental in elucidating the source of the discrepancies between experiment and theory and in verifying the surprising trend in the lifetimes. From the theory side, a development that would allow calculations of TDMs within the coupled cluster approach would be beneficial in this and in other important applications. We observe a sizeable discrepancy between our value for the lifetime of the $A'^2\Delta_{3/2}$ state of BaF, 5.3 $\mu$s, and the theoretical result from Ref. \cite{Chen2016s}, 220 ns. However, the latter value is an estimate based on the $A'^2\Delta_{3/2} - A^2\Pi_{3/2}$ mixing obtained from an effective Hamiltonian matrix, while our results comes from direct \textit{ab initio} calculations.

Finally, the products of transition rates $\Gamma_{n'v'n''v''}$ (\ref{rate}) with the corresponding lifetimes $\tau_{n'v'}$ (\ref{lifetime}) give the radiative branching ratios (relative decay fractions) shown in the Figure~\ref{fig:fractions}.

\begin{table}
\caption{Calculated lifetimes (ns) of the excited states of CaF, SrF, and BaF. \label{Lifetime}}
\begin{tabular}{c|lllrl}\hline \hline
\diagbox{Mol.}{State}& $A^2\Pi_{\frac{1}{2}}$ & $A^2\Pi_{\frac{3}{2}}$ & $B^2\Sigma^+_{\frac{1}{2}}$ & $A{^\prime} ^2\Delta_{\frac{3}{2}}$ & Ref.\\\hline
CaF&18.3&18.1&19.7&546&Present\\
&19.48$^e$&19.48$^e$&&&\cite{Pelegrini2005}\\
&21.9(4.0)$^a$&18.4(4.1)$^a$&25.1(4.0)$^a$&&Exp.\\

SrF&20.7&19.6&22.4&1130&Present\\
&24.1(2.0)$^a$&22.6(4.7)$^a$&25.5(0.5)$^c$& & Exp.\\

BaF&40.4&34.7&37.0&5289&Present\\
&37.8$^f$&37.8$^f$&&&\cite{Kang2016}\\
&&&&220&\cite{Chen2016s}\\
&56.0(0.9)$^d$&46.1(0.9)$^b$&41.7(0.3)$^b$&&Exp.\\
\hline\hline
\end{tabular}
\centering
\begin{tablenotes}                                                                                      \footnotesize
\item[\emph{a}]{$^a$Ref. \cite{Dagdigian1974}; $^b$Ref. \cite{Berg1993}; $^c$Ref. \cite{Berg1996}; $^d$Ref. \cite{Berg1998}; $^e$ This is derived from the calculated TDM using MRCI wave function; $^f$ This is derived from the calculated TDM with CASSCF+MRCI+SOC method for transition $A^2\Pi - X^2 \Sigma^+$}.
\end{tablenotes}

\end{table}

\section{Impact on laser-cooling}
\label{Lasercooling}

In this section we use the results of our molecular structure calculations to discuss the impact of laser-cooling applications for BaF molecules, and compare it to CaF and SrF, for which it has been demonstrated that laser-cooling works efficiently. Typically, scattering of a few thousand photons is sufficient to transversely cool heavy molecules in a molecular beam. In order to slow molecules from a buffer gas beam to below the capture velocity of a magneto-optical trap (MOT), tens of thousands of photons need to be scattered. The requirements for a MOT are even more stringent as the molecules need to continuously scatter photons to remain trapped (at a rate of typically 10$^6$ photons per second).

Molecules usually exhibit multiple decay paths from the excited state. The excited states under consideration here, the $A^2\Pi$ and the $B^2\Sigma^+$ states, have their lowest vibrational levels below the first dissociation limit, so that pre-dissociation is absent, and decay is purely radiative. As for rotation, the level structure is such that when exciting from an $N=1$ ground state level, in both $^2\Pi-^2\Sigma^+$ and $^2\Sigma^+ -^2\Sigma^+$ electronic transitions, an excited rotational level can be chosen that due to parity and angular momentum selection rules can only decay back to the $N=1$ ground state level (where it is assumed that rotational mixing due to external electric fields or due to nuclear spin can be neglected)~\cite{Stuhl2008}. Therefore, the main problem is decay to vibrationally excited levels in the ground state which are not governed by strict selection rules. 

\begin{figure*}[t]
\centering
\includegraphics[scale=1,width=1\linewidth]{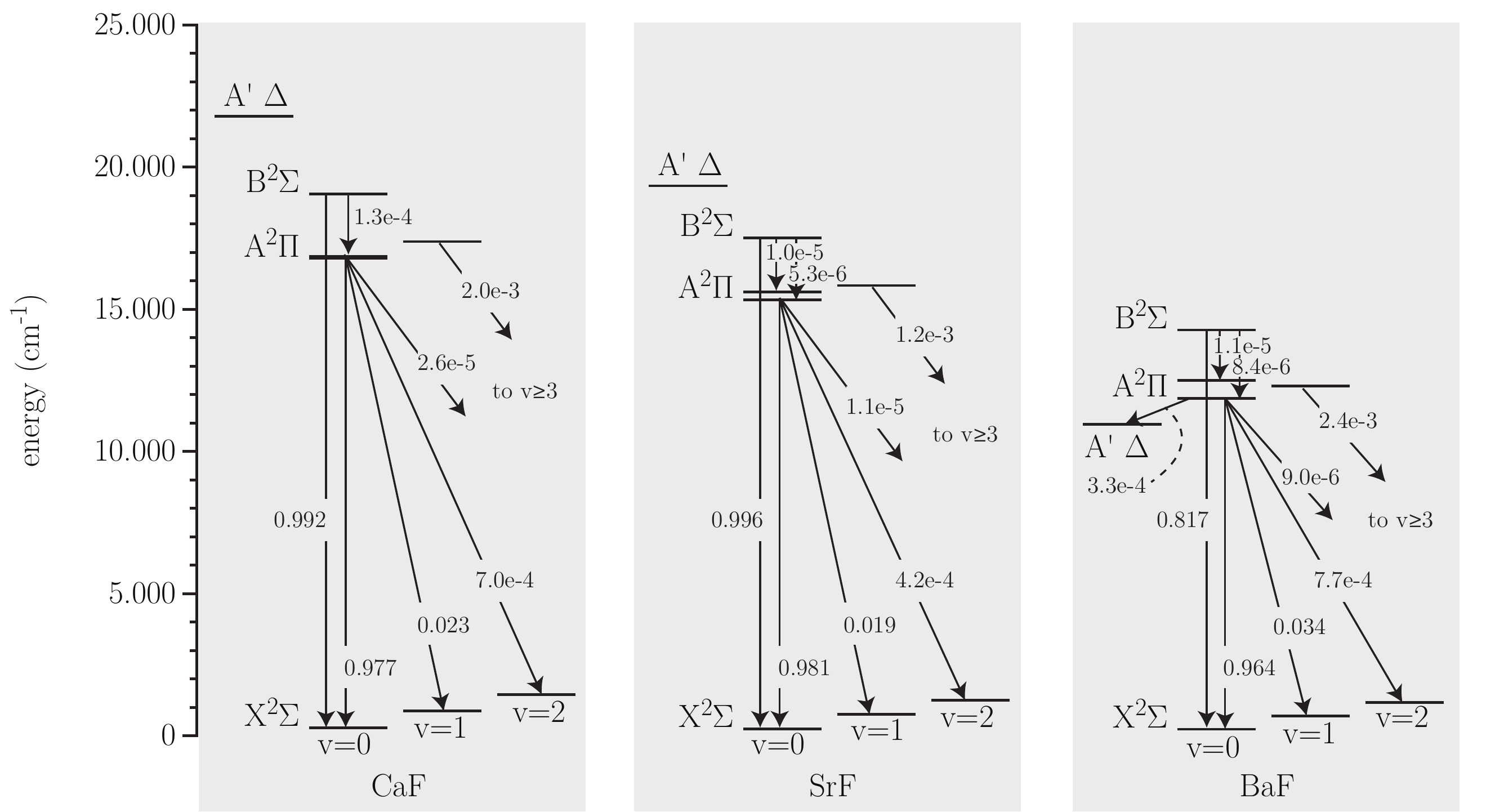}\\
\justifying 
\caption{The most important energy levels for laser-cooling and the calculated relative decay fractions for CaF, SrF, and BaF.}
\label{fig:fractions}
\end{figure*}

\begin{figure}[t]
\centering
\includegraphics[scale=1,width=1\linewidth]{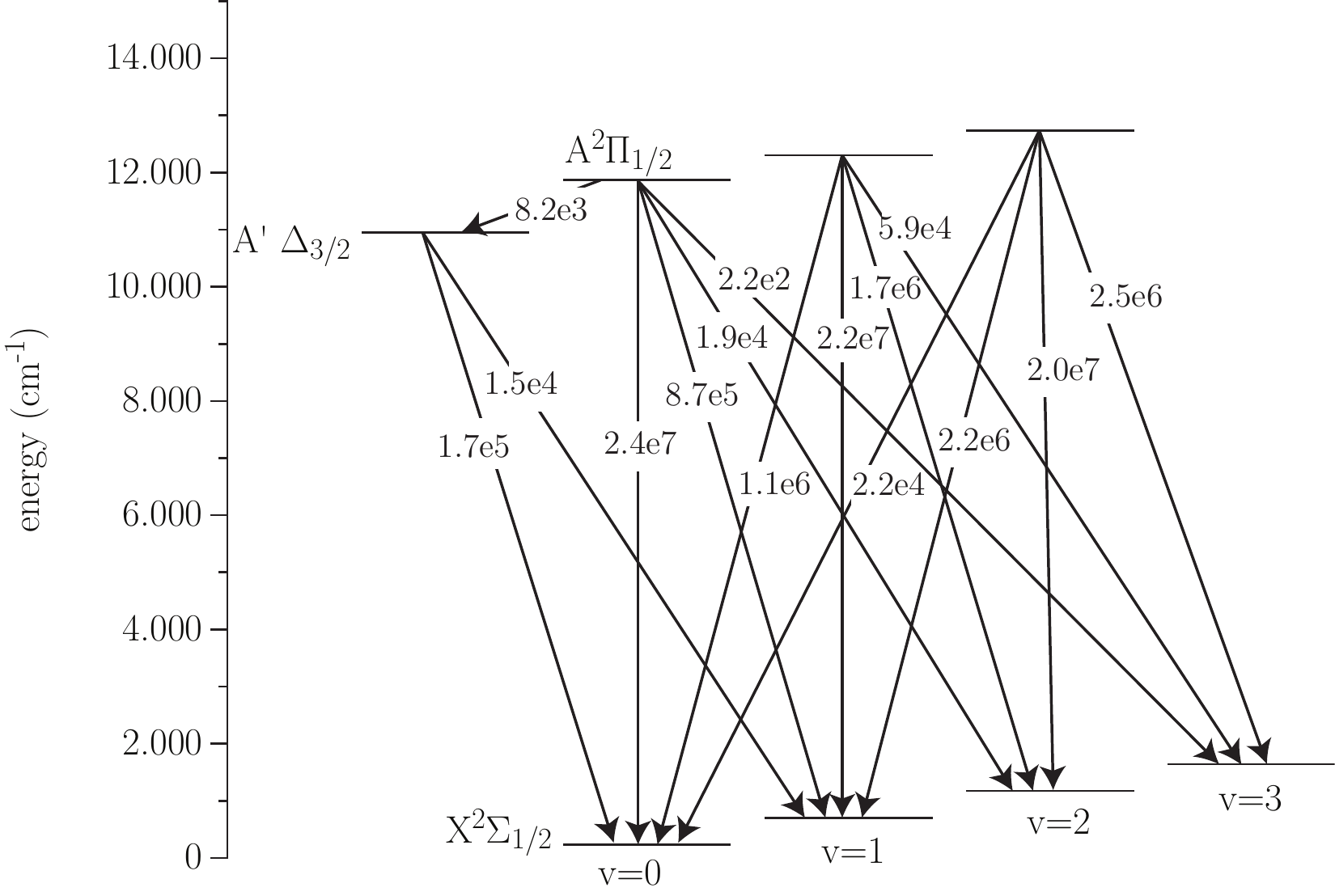}\\
\justifying 
\caption{Laser cooling level scheme of the $A^2 \Pi _{1/2} - X^2\Sigma^{+}_{1/2}$ system in BaF with the loss channel via the $A{^\prime}^2\Delta_{3/2}$ state. The absolute transition rates are given in units of s$^{-1}$.}
\label{fig:BaFrates}
\end{figure}

Based on the calculated absolute decay rates associated with the  lifetimes listed in Table~\ref{Lifetime}, relative decay fractions (branching ratios) have been calculated, taking into account decay from the $B^2\Sigma^+$ to the $A^2\Pi_{1/2}$ state and (for BaF) the decay from the $A^2\Pi$ state to the metastable $A{^\prime}^2\Delta$ state. The results are depicted in Fig.~\ref{fig:fractions}. Note that, in principle, laser-cooling via the $A^2\Pi_{3/2}$ state is also possible; however, the small $\Lambda$-splitting in this state would require reduction of the external electric fields to an impractically low level, and hence we will not consider this path further. From the relative decay fractions and the (experimental) transition frequencies, the transition rates have been calculated for the CaF, SrF, and BaF. For BaF, these are depicted in Figure~\ref{fig:BaFrates}.

In principle the procedure to find the optimal cooling scheme for each of these molecules is straightforward: start with the strongest transition with good Franck-Condon overlap, and add re-pump lasers to fix the leaks in order of importance. However, strong lasercooling via one excited state leads to an equal distribution of the molecules over all states involved which reduces the maximum optical scattering rate~\cite{Tarbutt2013}. Hence, for smaller leaks it is more attractive to use an alternative route back to the ground state~\cite{Smallman2014}.

\begin{table}[b]
\caption{Estimated number of photons scattered on a cycling transition before half of the molecules are lost. \label{tabel:scattered_photons}
}
\begin{ruledtabular}
\begin{tabular}{lllll}
Transition&Repump&CaF&SrF&BaF\\\hline
$X-A$&no repump &$29$&$36$&$19$\\
$ $& $v=1$ repump&$9.5 \times 10^2$&$1.6 \times 10^3$&$6.2 \times 10^2$\\
$ $& $v=2$ repump&$2.7 \times 10^4$&$6.2 \times 10^4$&$2.0 \times 10^3$\\
$ $&$\Delta$ repump& & &$7.2 \times 10^4$\\
$X-B$&no repump &$8.4 \times 10^2$ & $1.9 \times 10^2$&$3.4$\\
 & $v=1$ repump&$4.3 \times 10^3$ & $3.8 \times 10^4$&$42$\\
\end{tabular}
\end{ruledtabular}
\end{table}

Table~\ref{tabel:scattered_photons} lists the number of photons that can be scattered from CaF, SrF, and BaF via the $A^2\Pi_{1/2}$ and the $B^2\Sigma^+$ states, determined from the calculated transition rates. These numbers represent the maximum number of times that a given transition can be excited before on average half of the molecules will have decayed through a leak to another level. A number of observations can be made from this table. First of all, the large Franck-Condon factor (0.9992) of the $B-X$(0-0) in CaF allows one to scatter on average $8.4\times10^2$ photons before a molecule decays to an unwanted state. It should be noted that specifically this number is very sensitive to small deviations, since the FCF is so close to unity. According to our calculations, adding a re-pumper from the $v=1$ of the ground state gives only a limited increase as the decay from the $B$-state to the $A$-state is a significant loss channel. The Franck-Condon factors for the $A-X$ transition in CaF are somewhat less favorable, but still allow to scatter $2.7\times10^4$  photons using 2 lasers for re-pumping from the first and second vibrationally excited states of the ground state. The Franck-Condon factors of the $B-X$ transition of SrF are not as optimal as those of CaF, but the leak to the $A$ state is reduced. On the other hand, the Franck-Condon factors of the $A-X$ transitions are somewhat better than those of CaF, allowing to scatter on average $6.2\times10^4$ photons using 2 re-pumpers. Finally, the Franck-Condon factors of the $B-X$ transition of BaF are much smaller than those of CaF and SrF making laser-cooling on the $B-X$ transition impractical. The $A-X$ transition in BaF can be used to scatter $2.0\times10^3$ photons using 2 re-pumpers. Adding a laser to close the leak from the $v=1$ in the excited state to the $v=3$ in the ground state will not change much because decay to the $A{^\prime}^2\Delta$ state is a larger limiting factor. If one could close this leak to the $A{^\prime}^2\Delta$, the number of scattered photons would increase to $7.2\times10^4$. However, with an energy separation of $\sim$900 cm$^{-1}$ this is not straightforward technically. We conclude from this that although the $A-X$ transition is too leaky to be used for longitudinal slowing, sufficient photons can be scattered to perform transverse cooling. We note that due to its long lifetime, the $A{^\prime}^2\Delta$ state has a narrow linewidth. As a consequence, laser-cooling on the $X-\Delta$ transition may be used to reach a very low Doppler limit temperature~\cite{Collopy2015}.

\section{CONCLUSION\label{conclusion}}
The main goal of this work was investigation of the electronic structure of BaF, which will be used in an experiment to measure the electric dipole moment of the electron. Transverse laser-cooling of the BaF beam is an important component of the planned experiment, and knowledge of the internal structure of the molecule is necessary for identification of an efficient cooling scheme.

We present high-accuracy relativistic Fock space coupled cluster calculations of the potential energy curves and the spectroscopic constants of the ground and the lower excited states of the CaF, SrF, and BaF molecules. Our results for spectroscopic constants are in excellent agreement with experiment, where available, which gives credence to our predictions where no measurements were performed. Using the calculated potential energy curves, we obtain Franck-Condon factors for the $A^2\Pi_{1/2}-X ^2\Sigma^{+}_{1/2}$, $B^2\Sigma^{+}_{1/2}-X ^2\Sigma^{+}_{1/2}$, $A^2\Pi_{1/2}-A'^2\Delta_{3/2}$, and $A'^2\Delta_{3/2}-X ^2\Sigma^+_{1/2}$ transitions. The first two are possible cooling transitions that were previously successfully employed in laser-cooling of CaF and SrF. The investigation of the $A'^2\Delta_{3/2}$ state is due to the fact that it constitutes a potential leak in the BaF cooling cycle. We have also calculated the TDMs of these transitions, using relativistic multireference configuration interaction approach. Based on the calculated TDMs and experimental transition energies we determined the lifetimes of the excited states in BaF and its lighter homologues. The calculated FCFs and TDMs were also used to calculate the relative decay fractions and the transition rates for the three molecules. Finally, using the obtained molecular properties, we investigate the possible cooling schemes in BaF. The $B^2\Sigma^{+}_{1/2}-X ^2\Sigma^{+}_{1/2}$ cooling transition was shown to be extremely efficient in CaF; however, due to the non-diagonal nature of the FCFs for this transition in BaF, laser-cooling on this transition is impractical. The $A^2\Pi_{1/2}-X ^2\Sigma^{+}_{1/2}$ transition, on the other hand, seems much more promising. We have estimated that it is possible to scatter about 2000 photons on this transition (if two re-pump lasers are added to close the leaks to higher vibrational levels), which is sufficient for transverse laser-cooling. 

\section*{Acknowledgments}
The NL-$e$EDM consortium receives program funding from the Netherlands Organisation for Scientific Research (NWO). The authors thank the Center for Information Technology of the University of Groningen for support and for providing access to the Peregrine high performance computing cluster. AB acknowledges the University of Groningen for the Rosalind Franklin fellowship.  LFP wishes to acknowledge the support from the Slovak Research and Development Agency under the Contract No. APVV-15-0105. We thank Samir Tohme for insightful discussions.

\section*{Appendix \label{App}}

\begin{table*}[t]
\caption{\label{FCF_P32_XS12} Frank-Condon factors (FCFs) for vibronic transitions between the $\vert A^2\Pi_{\frac{3}{2}},v'\rangle$ and the $\vert X^2\Sigma^{+}_{\frac{1}{2}},v\rangle$ states of CaF, SrF, and BaF (present work, X2C-FSCC).\\}                                                                                                                                      
\begin{subtable}{1\textwidth}
\sisetup{table-format=-1.2}
\begin{ruledtabular}
\begin{tabular}{c|cccc}
\diagbox{$A^2\Pi_{\frac{3}{2}}$}{$X^2\Sigma^{+}_{\frac{1}{2}}$} & $v=0$ & $v=1$ & $v=2$ & $v=3$\\
\hline
${v}^{\prime}=0$ & $9.733\times10^{-1}$ & $2.574\times10^{-2}$ & $9.036\times10^{-4}$ & $3.745\times10^{-5}$\\         
${v}^{\prime}=1$ & $2.663\times10^{-2}$ & $9.220\times10^{-1}$ & $4.861\times10^{-2}$ & $2.564\times10^{-3}$\\
${v}^{\prime}=2$ & $4.763\times10^{-5}$ & $5.208\times10^{-2}$ & $8.738\times10^{-1}$ & $6.882\times10^{-2}$\\
${v}^{\prime}=3$ & $2.099\times10^{-7}$ & $1.347\times10^{-4}$ & $7.639\times10^{-2}$ & $8.286\times10^{-1}$\\
\end{tabular}
\end{ruledtabular}
\caption{CaF}\label{tab:FCF32}
\end{subtable}
\bigskip
\begin{subtable}{1\textwidth}
\sisetup{table-format=-1.2}    
\begin{ruledtabular}
\begin{tabular}{c|cccc}
\diagbox{$A^2\Pi_{\frac{3}{2}}$}{$X^2\Sigma^{+}_{\frac{1}{2}}$} & $v=0$ & $v=1$ & $v=2$ & $v=3$\\                                                                                                           
\hline

${v}^{\prime}=0$ & $9.769\times10^{-1}$ & $2.245\times10^{-2}$ & $5.955\times10^{-4}$ & $1.864\times10^{-5}$\\

${v}^{\prime}=1$ & $2.301\times10^{-2}$ & $9.320\times10^{-1}$ & $4.324\times10^{-2}$ & $1.728\times10^{-3}$\\

${v}^{\prime}=2$ & $6.225\times10^{-5}$ & $4.541\times10^{-2}$ & $8.886\times10^{-1}$ & $6.243\times10^{-2}$\\

${v}^{\prime}=3$ & $4.937\times10^{-9}$ & $1.850\times10^{-4}$ & $6.723\times10^{-2}$ & $8.468\times10^{-1}$\\
\end{tabular}
\end{ruledtabular}
\caption{SrF}\label{tab:FCF22}
\end{subtable}
\bigskip
\begin{subtable}{1\textwidth}
\sisetup{table-format=-1.2}    
\begin{ruledtabular}
\begin{tabular}{c|cccc}
\diagbox{$A^2\Pi_{\frac{3}{2}}$}{$X^2\Sigma^{+}_{\frac{1}{2}}$} & $v=0$ & $v=1$ & $v=2$ & $v=3$\\                                                                                           
\hline

${v}^{\prime}=0$ & $9.640\times10^{-1}$ & $3.515\times10^{-2}$ & $8.858\times10^{-4}$ & $1.122\times10^{-5}$\\

${v}^{\prime}=1$ & $3.554\times10^{-2}$ & $8.917\times10^{-1}$ & $6.994\times10^{-2}$ & $2.742\times10^{-3}$\\

${v}^{\prime}=2$ & $5.060\times10^{-4}$ & $7.144\times10^{-2}$ & $8.183\times10^{-1}$ & $1.040\times10^{-1}$\\

${v}^{\prime}=3$ & $9.061\times10^{-7}$ & $1.670\times10^{-3}$ & $1.072\times10^{-1}$ & $7.443\times10^{-1}$\\
\end{tabular}
\end{ruledtabular}
\caption{BaF}\label{tab:FCF12}
\end{subtable}
\end{table*}

\clearpage

\bibliography{BaF_total}

\end{document}